\documentclass[11pt]{amsart}
\usepackage{amssymb,latexsym,amsthm}
\usepackage{epsfig,graphicx, color}
\usepackage[mathscr]{euscript}
\usepackage[ansinew]{inputenc}

\newtheorem*{th*}{Theorem}
\newtheorem*{thm2bis}{Theorem $2'$}
\newtheorem*{thm3bis}{Theorem $3'$}

\DeclareMathOperator{\supp}{Supp}
\DeclareMathOperator{\per}{Per}
\DeclareMathOperator{\diam}{diam}
\DeclareMathOperator{\dist}{dist}
\def\div{{\text{div}}}

\def\Fl{{\mathscr F_\Lambda}}
\def\Pl{{\mathscr P_\Lambda}}
\def\mul{{\mu_\Lambda}}
\def\({\left(}
\def\){\right)}
\def\loc{{\text{\rm loc}}}

\def\indic{\mathbf{1}}

\def\Xint#1{\mathchoice
   {\XXint\displaystyle\textstyle{#1}}%
   {\XXint\textstyle\scriptstyle{#1}}%
   {\XXint\scriptstyle\scriptscriptstyle{#1}}%
   {\XXint\scriptscriptstyle\scriptscriptstyle{#1}}%
   \!\int}
\def\XXint#1#2#3{{\setbox0=\hbox{$#1{#2#3}{\int}$}
     \vcenter{\hbox{$#2#3$}}\kern-.5\wd0}}
\def\dashint{\Xint-}

\newdimen\AAdi%
\newbox\AAbo%
\def\AAk#1#2{\s_etbox\AAbo=\hbox{#2}\AAdi=\wd\AAbo\kern#1\AAdi{}}%
\def\AAr#1#2#3{\s_etbox\AAbo=\hbox{#2}\AAdi=\ht\AAbo\raise#1\AAdi\hbox{#3}}%
\font\tenmsb=msbm10 at 12pt
\font\sevenmsb=msbm7 at 8pt
\font\fivemsb=msbm5 at 6pt
\newfam\msbfam
\textfont\msbfam=\tenmsb
\scriptfont\msbfam=\sevenmsb
\scriptscriptfont\msbfam=\fivemsb
\def\Bbb#1{{\mathbb#1}}

\newcommand{\pr}{\begin{proof}}
\newcommand{\cqfd}{\end{proof}}
\newcommand{\beq}{\begin{equation}}
\newcommand{\eeq}{\end{equation}}
\newcommand{\beqr}{\begin{eqnarray}}
\newcommand{\eeqr}{\end{eqnarray}}
\newcommand{\ba}{\begin{array}}
\newcommand{\ea}{\end{array}}

\begin{document}

\newtheorem{thm}{Theorem}
\newtheorem{lem}{Lemma}
\newtheorem{cor}{Corollary}
\newtheorem{rem}{Remark}
\newtheorem{pro}{Proposition}
\newtheorem{defi}{Definition}
\newtheorem{conj}[thm]{Conjecture}
\newcommand{\noi}{\noindent}
\newcommand{\dis}{\displaystyle}
\newcommand{\mint}{-\!\!\!\!\!\!\int}
\newcommand{\avint}{\mathop{\ooalign{$\displaystyle \int$\cr$-$}}}

\def \bx{\hspace{2.5mm}\rule{2.5mm}{2.5mm}} \def \vs{\vspace*{0.2cm}}
\def\hs{\hspace*{0.6cm}}
\def \ds{\displaystyle}
\def \p{\partial}
\def \O{\Omega}
\def \o{\omega}
\def \b{\beta}
\def \m{\mu}
\def \l{\lambda}
\def\L{\Lambda}
\def \ul{u_\lambda}
\def \D{\Delta}
\def \d{\delta}
\def \k{\kappa}
\def \s{\sigma}
\def \e{\varepsilon}
\def \a{\alpha}
\def \tf{\tilde{f}}
\def \rmk {\noindent {\it Remark} }
\def \esp {\hspace{4mm}}
\def \dsp {\hspace{2mm}}
\def \ssp {\hspace{1mm}}

\def \u{u_+^{p^*}}
\def \ui{(u_+)^{p^*+1}}
\def \ul{(u^k)_+^{p^*}}
\def \energy{\int_{\R^n}\u }
\def \sk{\s_k}
\def \mo{\mu_k}
\def\cal{\mathcal}
\def \I{{\cal I}}
\def \J{{\cal J}}
\def \K{{\cal K}}
\def \OM{\overline{M}}

\def\fk{{{\cal F}}_k}
\def\M1{{{\cal M}}_1}
\def\Fk{{\cal F}_k}
\def\FF{\cal F}
\def\Gk{{\Lambda_k^+}}
\def\n{\nabla}
\def\uuu{{\n ^2 u+du\otimes du-\frac {|\n u|^2} 2 g_0+S_{g_0}}}
\def\uuug{{\n ^2 u+du\otimes du-\frac {|\n u|^2} 2 g+S_{g}}}
\def\sku{\sk\left(\uuu\right)}
\def\vvv{{\frac{\n ^2 v} v -\frac {|\n v|^2} {2v^2} g_0+S_{g_0}}}
\def\vvs{{\frac{\n ^2 \tilde v} {\tilde v}
 -\frac {|\n \tilde v|^2} {2\tilde v^2} g_{S^n}+S_{g_{S^n}}}}
\def\skv{\sk\left(\vvv\right)}
\def\tr{\hbox{tr}}
\def\pO{\partial \Omega}
\def\dist{\hbox{dist}}
\def\RR{\Bbb R}\def\R{\Bbb R}
\def\C{\Bbb C}
\def\B{\Bbb B}
\def\N{\Bbb N}
\def\Q{\Bbb Q}
\def\Z{\Bbb Z}
\def\PP{\Bbb P}
\def\EE{\Bbb E}
\def\F{\Bbb F}
\def\G{\Bbb G}
\def\H{\Bbb H}
\def\SS{\Bbb S}\def\S{\Bbb S}
\def\L2{{L^2_\loc(\R^2\setminus\Lambda,\R^2)}}

\def\circledwedge{\setbox0=\hbox{$\bigcirc$}\relax \mathbin {\hbox
to0pt{\raise.5pt\hbox to\wd0{\hfil $\wedge$\hfil}\hss}\box0 }}

\def\sss{\frac{\s_2}{\s_1}}

\date{}
\title[Renormalized energy] {On lattices with finite Coulombian interaction energy in the plane}

\author{Yuxin Ge}
\address{Laboratoire d'Analyse et de Math\'ematiques Appliqu\'ees,
CNRS UMR 8050,
D\'epartement de Math\'ematiques,
Universit\'e Paris Est-Cr\'eteil Val de Marne \\61 avenue du G\'en\'eral de Gaulle,
94010 Cr\'eteil Cedex, France}
\email{ge@u-pec.fr}
\author{Etienne Sandier}
\address{Laboratoire d'Analyse et de Math\'ematiques Appliqu\'ees,
CNRS UMR 8050,
D\'epartement de Math\'ematiques,
Universit\'e Paris Est-Cr\'eteil Val de Marne \\61 avenue du G\'en\'eral de Gaulle,
94010 Cr\'eteil Cedex, France}
\email{sandier@u-pec.fr}
\begin{abstract} We present criteria for the Coulombian interaction energy of infinitely many points in the plane with a uniformly charged backgroud introduced in \cite{ss1} to be  finite, as well as examples. We also show that in this unbounded setting, it is not always possible to project an $L^2_\loc$ vector field onto the set of gradients in a way that reduces its average $L^2$ norm on large balls. 
\end{abstract}

\maketitle
\section{Introduction}
Given a discrete set $\Lambda$  in the plane (we will also say a {\em lattice}) and a real number $m\ge 0$, the renormalized  energy introduced in 
\cite{ss1} heuristically describes the interaction energy of unit charges placed at the points of $\Lambda$ with a uniform negative background of density $m\in\R$. It is defined in several steps, following mostly \cite{ss1}.  

First, denoting $\nu:=\sum_{p\in\Lambda}\delta_{p}$  and for any  vector-field  $j$ solving 
\begin{equation}
\label{equation1}
-\div(j) =2\pi(\nu-m)\mbox{ in
}\R^2,
\end{equation}
and belonging to $\L2$ we define $W(j)$ as follows: For any $R>1$ we denote by $\chi_R$ a smooth approximation of the indicator function of $B_R$, the ball centered at $0$ with radius $R$. More precisely we assume that 
\beq\label{chir}
\text{$\chi_R\ge 0$, $\|\nabla \chi_R\|_\infty\le C$, $\chi_R\equiv 1$ on $B_{R-1}$ and $\chi_R\equiv 0$ on $\R^2\setminus B_R$},
\eeq
where $C$ is independent of $R$. Then we let 
\begin{equation}\label{energyrenormalized1}
W(j):=\limsup_{R\to\infty}\frac {W(j,\chi_R)}{|B_R|},\quad 
W(j,\chi_R):=\limsup_{\eta\to 0}\frac12\int_{\R^2\setminus\cup_{p\in \Lambda
}B(p,\eta)}\chi_R |j|^2+\pi\log\eta\sum_{p\in\Lambda}\chi_R(p )
\end{equation}

Second, we consider  the set $\Fl$ of vector fields in $\L2$ satisfying \eqref{equation1} for a given $\Lambda$ and $m$, and the subset $\Pl$ of curl-free vector fields in $\Fl$, or equivalently the set of those elements in $\Fl$ which are gradients. We may now define 
\begin{equation}\label{energyrenormalized2}
W(\Lambda):= \inf_{\nabla U\in \Pl} W(\nabla U), \quad 
\tilde W(\Lambda):=\inf_{j\in \Fl} W(j).
\end{equation}
Note that  $\Fl$ and $\Pl$ depend on $m$, hence so do $W(\Lambda)$ and $\tilde W(\Lambda)$. But in fact (see below) the value of $m$ is determined by $\Lambda$ in the sense that $W(\Lambda)$ or $\tilde W(\Lambda)$ can only be finite for at most one value of $m$ (which is the asymptotic density of $\Lambda$ whenever it exists). In any case, the value of $m$ will always be clear from the context or made precise. 

\begin{rem}
It will be useful to generalize somewhat the above definition to allow $j$'s satisfying  \eqref{equation1} with 
$$\nu:=\sum_{p\in\Lambda}\alpha_p\delta_{p}.$$
In this case one should modifiy the definition of $W(j,\chi_R)$: 
\begin{equation}\label{weight}
W(j,\chi_R):=\limsup_{\eta\to 0}\frac12\int_{\R^2\setminus\cup_{p\in \Lambda
}B(p,\eta)}\chi_R |j|^2+\pi|\alpha_p|^2\log\eta\sum_{p\in\Lambda}\chi_R(p ).
\end{equation}
\end{rem}

In \cite{ss1}, only $W$ is considered. One could think at first that $W$ and $\tilde W$ are equal, the argument being the following: Since  $W(j)$ may be seen as the average of $|j|^2$ over $\R^2$ (with the infinite part due to the Dirac masses in \eqref{equation1} removed), then projecting onto the set of curl-free fields would reduce this quantity, so that the infimum of $W(j)$ over $\Fl$ would in fact be acheived by some $j\in\Pl$, proving that $W(\Lambda) = \tilde W(\Lambda)$. It turns out however that this is not the case and in fact we prove (see Theorem~\ref{thm1} below) that with $m=0$, 

\begin{th*}\label{intro1} $W(\N) = +\infty$ and $\tilde W(\N) < +\infty$.\end{th*}

The rest of the paper is devoted to giving sufficient conditions on $\Lambda$ for $\tilde W$ and/or $W$ to be finite. There are roughly two factors which can make $W$ or $\tilde W$ infinite. First, there is the logarithmic interaction between pairs of points, which can be made infinite by bringing points very close to each other: we will not consider this factor here and to rule it out we restrict ourself to {\em uniform} $\Lambda$'s in the following sense.
\begin{defi} Given a lattice $\Lambda$  and weights  $\{\alpha_p\}_{p\in\Lambda}$, we say  that $$\nu = 2\pi\sum_{p\in \Lambda}\alpha_p\delta_p$$
is of {\em uniform type} if 
$$\min_{p\neq q\in \Lambda}|p-q|>0,\quad \sup_{p\in\Lambda}|\alpha_p|<\infty.$$
If the weights are all equal to $1$ we simply say $\Lambda$ is of uniform type.
\end{defi}

The second factor which can make $W$ or $\tilde W$ infinite is the interaction with the background. If we restrict ourselves to uniform $\Lambda$'s, then for a given $m$ the quantities $W(\Lambda)$ or $\tilde W(\Lambda)$ measure how close $\sum_{p\in \Lambda}\delta_p$ is to a uniform density $m$. Our second main result shows that this can be measured by simply counting the number of points of $\Lambda$ in any given ball (see Theorems~\ref{thm2} and \ref{thm2bis}). In particular we have 
\begin{th*}\label{intro2}
Assume  that $\Lambda$ is uniform and that  there exists $m, C\ge 0$ and  $\e\in (0,1)$  such that for any $x\in\R^2$ and  $R>1$ we have, denoting $\sharp E$ the number of elements in $E$, 
\begin{equation}\label{bound}
\left|\sharp\( B(x,R)\cap \Lambda\)-m\pi R^2\right|\le CR^{1-\e}
\end{equation}
Then $W(\Lambda)<+\infty$ for this value of $m$.
\end{th*}

This criterion for finiteness is optimal in the sense that if we replace the right-hand side in \eqref{bound} by $C R^{1+\e}$, then it is not difficult to construct $\Lambda$'s satisfying \eqref{bound} and having infinite renormalized energies (see Proposition~\ref{opt}). This criterion can be relaxed a bit in the case of $\tilde W$ (see Theorem~\ref{thm2bis}).

This leaves open the case $\e = 0$ (in which case $\N$ and $\Z$ satisfy \eqref{bound} with $m=0$). In this case we are able to prove a partial result (see Theorem~\ref{criticalbis} for a variant)
\begin{th*}
\label{intro3}
Let $A\subset \Z^2$ and $\Lambda:= \Z^2\setminus A$. Assume there exists some constant $C>0$ such that for all $x\in \R^2$ and $R>1$ we have
$$
\sharp\(A\cap B(x,R)\)\le CR.
$$
Then $\tilde W(\Lambda)<+\infty.$
\end{th*}
The proof of this theorem is based on the fact (see Proposition~\ref{procritical}) that under the above hypothesis there exists a bijection between $\Z^2\setminus A$ and $\Z^2$ under which points are moved at uniformly bounded distances. This is a discrete analogue of a result of G.Strang \cite{strang}. 

The criterion in Theorem~\ref{intro2} is satisfied by perfect (or Bravais) lattices, or more generally by doubly periodic lattices  (see \cite{reseaux}) --- even though in this case (see below) the conclusion of Theorem~\ref{intro2} is almost trivial. However we are not aware that this is known for quasiperiodic lattices, and thus  we give a construction similar to that of Theorem~\ref{intro2} which allows us to conclude for an exemple of Penrose-type lattice $\Lambda$ that $\tilde W(\Lambda)<+\infty$. We have not sought generality in this direction, and refer to  Section~\ref{penrose} for the construction of $\Lambda$ and the proof that $\tilde W(\Lambda)$ is finite. 

\section{Some properties of $W$, $\tilde W$}
We always assume the following property of $\nu := \sum_{p\in\Lambda} \delta_{p}$, which is satisfied in particular if $\Lambda$ is uniform. 
\begin{equation}
\label{equation2}
\limsup_{R\to+\infty}\frac{\nu(B_R)}{|B_R|}<+\infty.
\end{equation}
We begin by recalling some facts from \cite{ss1, ss2}. 
\begin{description}
\item[Structure of $\Pl$] If $\Lambda$ satisfies \eqref{equation2} and $W(\Lambda)$ is finite, then the  set $\{\nabla U\in \Pl\mid W(\nabla U)<+\infty\}$ is a 2-dimensional affine space. Any  two gradients in this set differ by a constant vector.  
\item[Minimization] For any given $m$, the function $\Lambda\to W$ defined over the set of $\Lambda$'s satisfying \eqref{equation2} is bounded from below and admits a minimizer. 
\item[Scaling] Denote $W_m$ the renormalized energy with background $m\in\R$. If $j$ satisfies \eqref{equation1} and \eqref{equation2} holds, then 
  \[
 W(j)=m\(W(j')-\frac{\pi}{2}\log m\),\quad\text{with}\quad j'(\cdot)=\frac{1}{\sqrt{m}}j\(\frac\cdot {\sqrt{m}}\).
  \]
\item[Cutoffs] If \eqref{equation2} holds, then the value of  $W(\Lambda)$ (or $\tilde W(\Lambda)$)    does not depend on the particular choice of cut-off functions $\chi_R$ as long as they satisfy the stated properties. 
\item[Perfect lattices] Assume $\Lambda=\Z \vec{u}\oplus\Z \vec{v}$ where $(\vec{u},\vec{v})$ is a basis of $\R^2$ satisfying the normalized volume condition $|\vec{u}\wedge \vec{v}|=1$. Let $\Lambda^*$ be the dual lattice of $\Lambda$. Then, taking $m=1$, 
$$
W(\Lambda)=\pi \lim_{x\to 0}\left(\sum_{p\in \Lambda^*\setminus \{0\}} \frac{e^{2i\pi p\cdot x}}{4\pi^2|p|^2}+\log|x|\right)-\frac\pi 2\log2\pi.
$$
Moreover, the minimum of $W$ among lattices of this type is acheived by the triangular lattice
$$
\Lambda_1:=\sqrt{\frac{2}{\sqrt{3}}}\left((1,0)\Z \oplus \left(\frac12,\frac{\sqrt{3}}{2}\right)\Z\right).
$$
\item[Uniqueness of $m$] For a given  $\Lambda$, there can be at most one value of $m$ for which $W(\Lambda)<+\infty$. Indeed if $j_1$ (resp. $j_2$) satisfy \eqref{equation1} with $m_1$ (resp. $m_2$) then $-\div(j_1-j_2) = m_2 - m_1$, and  if $m_1\neq m_2$ this implies that $W(j_1)$ and $W(j_2)$ cannot both be finite.  To see this one can use Proposition~\ref{propprel} below in case the points in $\Lambda$ are uniformly spaced. Otherwise one has to resort to the corresponding result in \cite{ss1}. 
\end{description}

One of the main points in \cite{ss1,ss2} is the fact that $W$ is bounded below. This is in fact very easy to prove in the case of $\Lambda$'s --- or more generally $\nu$'s --- which are of uniform type. It is a consequence of the following useful fact.

\begin{pro}\label{propprel} If  $j$ satisfies \eqref{equation1} with $\nu$  of uniform type, then  for any $\delta<\frac12 \inf_{p\neq q\in \Lambda}|p-q|$ there exists $g:\R^2\to\R$ and $C>0$ such that 
\beq
  \label{lowbound}
 g\ge -C, 
 \eeq
 such that 
\beq
 \label{outside} g=\frac12 |j|^2,\quad \text{on}\quad \R^2\setminus \cup_{p\in \Lambda }B(p,\delta),
 \eeq
and such that for any compactly supported lipschitz function $\chi$, 
 \beq
 \label{error}
 \left|\int_{\R^2}\chi g-W(j,\chi)\right|\le CN\|\nabla\chi\|_\infty,
\eeq
where $N = \sharp\{p\in\Lambda\mid B(p,\delta)\cap \supp\nabla\chi\neq\varnothing\}$.
\end{pro}

\begin{rem} Note that if we take $\chi$ such that  $\chi = 1$ on $B(p,\delta)$ and $\chi = 0$ on every other $B(q,\delta)$ for $q\neq p\in\Lambda$ then \eqref{error} implies that $\int_{\R^2}\chi g = W(j,\chi)$. This implies in particular, approximating the indicator function $\indic_{B(p,\delta)}$ by such functions, that for any $p\in\Lambda$
\beq\label{indic} \int_{B(p,\delta)} g = W(j,\indic_{B(p,\delta)})\eeq
\end{rem}

\pr In $\R^2\setminus \cup_{p\in \Lambda } B(p,\delta)$, we let $g=\frac12 |j|^2$. Then, for any $p\in\Lambda$ and any $r\in (0,\delta)$ such that  $|j|\in L^2(\p B(p,r))$ --- this is the case for a.e. $r$ --- we define $\lambda_{p,r}>0$ to be a value of $\lambda$ such that  
 \beq
 \label{circle}
\frac 12 \int_{\p B(p,r)} \min(|j|^2,\lambda)=\frac{\pi\alpha_p^2}{r}  - 2\pi^2\alpha_p m r.
\eeq
The fact that $\lambda_{p,r}$ is well defined follows from the fact that the left-hand side of \eqref{circle} is a continuous increasing function of $\lambda$ which increases from $0$ to  (as $\lambda\to +\infty$)
$$\frac 12 \int_{\p B(p,r)} |j|^2 \ge \frac1{4\pi r}\(\int_{\p B(p,r)} j\cdot\nu\)^2 = \frac\pi r\(\alpha_p - m\pi r^2\)^2\ge \frac{\pi\alpha_p^2}{r}- 2\pi^2\alpha_p m r.$$

For any $r\le\delta$ we let, on $\partial B(p,r)$, 
$$
g:=\frac12(|j|^2- \lambda_{p,r})_+  - \pi\alpha_p m - \frac{\alpha_p^2}{\delta^2}\log\frac1{\delta}.
$$
Then \eqref{outside} is obviously satisfied, and \eqref{lowbound} is satisfied with $$C = \(\sup_{p\in\Lambda} {\alpha_p}\)^2 \frac{|\log\delta|}{\delta^2} + \pi |m| \sup_{p\in\Lambda}|\alpha_p|.$$ 
It remains  to prove \eqref{error}. For any function $\chi$ and any $\eta\le \delta$ we have 
\beq\label{aa}\ds\int_{\R^2\setminus \cup_{p\in \Lambda } B(p,\eta)} \chi \(\frac{ |j|^2}{ 2} - g\)=\ds\sum_{p\in\Lambda}\int_{B(p,\delta)\setminus B(p,\eta)}\chi  \(\frac{ |j|^2}{ 2}- g\).
\eeq
Then, writing $A$ for the annulus $B(p,\delta)\setminus B(p,\eta)$, 
\beq\label{bb}\int_{A}\chi \(\frac{ |j|^2}{ 2} - g\) = \chi(p)\int_{A}\(\frac{ |j|^2}{ 2}-g \) + \int_{A}(\chi -\chi (p ))\(\frac{ |j|^2}{ 2}-g\). \eeq
We have for any $r\le\delta$, on $\partial B(p,r)$
$$\(\frac{ |j|^2}{ 2}-\frac12(|j|^2- \lambda_{p,r})_+\) =\frac12 \min\( |j|^2,\lambda_{p,r}\),$$
hence using \eqref{circle} we find 
\begin{multline}\label{cc} \int_{A}\(\frac{ |j|^2}{ 2}-g\)= \int_\eta^\delta\frac{\pi\alpha_p^2}{r}- 2\pi^2\alpha_p m r\,dr  + \pi (\delta^2 - \eta^2) \( \pi\alpha_p m +\frac{\alpha_p^2}{\delta^2}\log\frac1{\delta}\) \\= \pi\alpha_p^2\log\frac1\eta -\pi\eta^2\frac{\alpha_p^2}{\delta^2}\log\frac1{\delta}.\end{multline}
On the other hand, since $|\chi -\chi (p )|\le r \|\nabla\chi\|_\infty$, and using \eqref{circle} we have  
$$\left|\int_{A}(\chi -\chi (p ))\(\frac{ |j|^2}{ 2}-g\)\right|
\le \|\nabla\chi\|_\infty\left|\int_\eta^\delta \frac r2\left(\left|\int_{\partial B(p,r)}\min\( |j|^2,\lambda_{p,r}\)\right|+Cr\right)dr\right|\le C \|\nabla\chi\|_\infty.$$
This together with \eqref{aa},\eqref{bb} and \eqref{cc} yields
$$\left|\lim_{\eta\to 0}\int_{\R^2\setminus \cup_{p\in \Lambda } B(p,\eta)} \chi \(g-\frac{ |j|^2}{ 2}\)- \sum_{p\in\Lambda }\pi\chi( p) \alpha_p^2\log \eta\right|\le C N\|\nabla\chi\|_\infty,$$
where $N = \sharp\{p\in\Lambda\mid B(p,\delta)\cap \supp\nabla\chi\neq\varnothing\}$. This proves \eqref{error}.
\cqfd

Note that, contrary to the corresponding result in \cite{ss1}, we have not proved that the constant $C$  in \eqref{lowbound} is universal, which\ is a delicate point. We have included this weaker result for the sake of self-containedness and because it has a simple proof. 


\section{Examples of finite or infinite energy lattices.}

We begin by showing that moving the points in $\Z^2$ at a bounded distance yields a lattice $\Lambda$ with finite energy, assuming $\Lambda$ is uniform.  
\begin{pro} 
\label{pro-move}
Let $\Lambda$ be a  lattice in the plane  satisfying $\inf_{x,y\in\Lambda, x\neq y}|x-y|>0$ and let $\Phi: \Lambda\to \Z^2$ be a bijective map such that  $\sup_{p\in \Lambda}|\Phi(p ) -p|<\infty$. Then $\tilde W(\Lambda)<+\infty$, with $m=1$.
\end{pro}
\pr
Let $R_1 = 2 \sup_{p\in \Lambda}|\Phi(p ) -p|$. Then for every $p\in \Lambda$, we solve
\[
\left\{
\begin{array}{lllll}
-\triangle U_p &=&\ds2\pi\left(\delta_{p}-\delta_{\Phi(p)} \right)&\mbox{ in
}B(p,R_1) \\
 \frac{\p U_p}{\p \nu}&=&0& \mbox{ on }\p B(p,R_1) 
\end{array}
\right.
\]
where $\nu$ is the outer unit normal on the boundary. Let $V$ be the $\Z^2$-periodic solution --- which is unique modulo an additive constant --- of 
\[
\begin{array}{lllll}
-\triangle V&=&\ds2\pi\left(\sum_{p\in \Z^2}\delta_{p}-1\right)&\mbox{ in
}\R^2
\end{array}
\]
Then by periodicity $|V(x)+\log|x-p||$ is bounded in  $C^2\(\cup_{p\in \Z^2} B(p,1/4)\)$, while $V(x)$  is bounded in $C^2$ of the complement.  More precisely we have the (see for instance \cite{ss1}) 
$$
V(x)=\sum_{p\in \Z^2\setminus \{0\}}\frac{e^{2i\pi p\cdot x}}{2\pi|p|^2}
$$
Now we  define $j:\R^2\to \R$ by
$$
j=\nabla V+\sum_{p\in \Lambda} \nabla U_p, 
$$
where $\nabla U_p$ is is extended by $0$ outside of $B(p,R_1) $ and thus defined on the whole of $\R^2$. 

From the assumptions on $\Lambda$ and $\Phi$ the sum above is finite on any compact set and thus $j$ is well defined and  solves 
 \[
\begin{array}{lllll}
-\div(j)&=&\ds2\pi\left(\sum_{p\in \Lambda}\delta_{p}-1\right)&\mbox{ in
}\R^2.
\end{array}
\]

On the other hand, $U_p(x)+\log |x-p|-\log|x- \Phi(p)|$ is bounded in $C^2(B(p,R_1))$, uniformly with respect to $p\in\Lambda$.  It follows that $j+\nabla \log|x- p|$ is bounded in $B(p,\delta)$ uniformly with respect to $p\in\Lambda$, and $j$ is bounded in $\R^2\setminus \cup_{p\in \Lambda} B(p,\delta)$, where $\delta>0$ is half the minimal distance between points of $\Lambda$. A straightforward consequence is that $W(j)<+\infty$ and then $\tilde W(\Lambda)<+\infty$.\cqfd

We will prove below that the conclusion in the above proposition cannot be improved to $W(\Lambda)<+\infty$. 

A  consequence of Proposition~\ref{pro-move} is 
\begin{cor} 
\label{cor1}
We have 
$$
\tilde W(\Z^2\setminus \Z)<\infty,\;\tilde W(\Z^2\setminus \N)<\infty
$$
 \end{cor}
 \pr
 We construct a bijective map from $\Phi:\Z^2\setminus \Z\to \Z^2$ by 
 $$
 \Phi(p_1,p_2)=\left\{
 \begin{array}{lll}
 (p_1, p_2-1)&\mbox{ if }p_2\ge 1\\
  (p_1, p_2)&\mbox{ if }p_2< 0
 \end{array}
 \right.
 $$
 The desired result follows from the above proposition. The proof for  $\Z^2\setminus \N$ is similar.
 \cqfd

A second tool for constructing $j$'s with finite energy is 
\begin{pro}
\label{prop2}
Assume $j_1$ (resp. $j_2$) satisfy  \eqref{equation1} with a $\nu_1$ (resp. $\nu_2$) of uniform type. Assume also that $\nu_1$ and $\nu_2$ satisfy \eqref{equation2} and that $\nu_1+\nu_2$ is of uniform type.

Then, if $W(j_1)<\infty$ for a background $m_1$ and $W(j_2)<\infty$ for the background $m_2$, we have 
$$
\text{$W(j_1+j_2)<\infty$ for the background $m_1+m_2$.}
$$
\end{pro}

First, we prove two lemmas.\\

\begin{lem}
\label{lem1}
Assume $j$ satisfies \eqref{equation1} and \eqref{equation2} with $\nu$  of uniform type, and assume  $W(j)<\infty$. Then there exists some positive constant $C$ depending  on $j$ such that for any $R>1$ and $\delta < \frac12 \inf\{|p-q|\mid p\neq q\in\Lambda\}$,
$$
\int_{B_R\setminus\cup_{p\in \Lambda}B(p,\delta)}|j|^2\le CR^2,\;\;\int_{B_R\cap\cup_{p\in \Lambda}B(p,\delta)}|j-G|^2\le CR^2,
$$
where $\ds G(x):=\alpha_p\frac{x-p}{|x-p|^2}$ if $x\in B(p,\delta)$ with $p\in \Lambda$.
\end{lem}
\pr
Let $g$ be constructed in Proposition \ref{propprel}.  From (\ref{error}), we have 
$$
\int \chi_Rg \le W(j,\chi_R) + C n( R )\le CR^2, 
$$
where $n(R ):=\sharp(\Lambda\cap B_{R+1})$. Hence 
\beq
\label{eq2.1.1}
 \int \chi_R g\le CR^2.
\eeq
On the other hand, since $g\ge -C$ and from the properties of $\chi_R$, we have  
\beq
\label{eq2.1.2}
 \int \chi_R g\ge \int_{B_R}g-CR\ge \int_{B_R\setminus\cup_{p\in \Lambda}B(p,\delta)}\frac12|j|^2+\sum_{p\in \Lambda,B(p,\delta)\subset B_R}\int_{B(p,\delta)}g-CR.
\eeq
For any $p\in \Lambda$, we define 
$$
\begin{array}{llll}
\ds W(j,\indic_{B(p,\delta)}):=\limsup_{\eta\to 0}\frac12\int_{B(p,\delta)\setminus B(p,\eta)}|j|^2+\pi \alpha_p^2\log\eta 
\end{array}
$$\
We have, denoting $A = B(p,\delta)\setminus B(p,\eta)$, 
\[
\begin{array}{llll}
\ds \frac12\int_{A}|j|^2&=&\ds\frac12\int_{A}|G|^2+|j-G|^2+2G\cdot(j-G)\\\
&=&\ds\pi \alpha_p^2\log\frac{\delta}{\eta}+\frac12\int_{A}|j-G|^2+\alpha_p\int_\eta^\delta \frac{dr}{r} \int_{\p B(p,r)} \nu\cdot (j-G) \\
&=&\ds\pi \alpha_p^2\log\frac{\delta}{\eta}+\frac12\int_{A}|j-G|^2+\alpha_p\int_\eta^\delta \frac{dr}{r}\int_{ B(p,r)}\div(j-G)\\
&=&\ds\pi \alpha_p^2\log\frac{\delta}{\eta}+\frac12\int_{A}|j-G|^2+\pi^2 \alpha_p m(\delta^2-\eta^2).
\end{array}
\]
Hence, we obtain 
\beq
\label{eq2.1.3.0}
W(j,\indic_{B(p,\delta)}) = \limsup_{\eta\to 0}\frac12\int_{A}|j|^2+\pi \alpha_p^2\log\eta= \pi \alpha_p^2\log \delta+\frac12\int_{B(p,\delta)}|j-G|^2+\pi^2 \alpha_p m\delta^2
\eeq
Thus, using \eqref{indic}, 
\beq \label{eq2.1.3}
\int_{B(p,\delta)}g=   \pi \alpha_p^2\log \delta+\frac12\int_{B(p,\delta)}|j-G|^2+\pi^2 \alpha_p m\delta^2
\eeq
Gathering (\ref{eq2.1.1}) to  (\ref{eq2.1.3}), we get
$$
CR^2\ge  \int \chi_R g \ge \int_{B_R\setminus\cup_{p\in \Lambda}B(p,\delta)}\frac12|j|^2+\sum_{p\in \Lambda,B(p,\delta)\subset B_R}\frac12\int_{B(p,\delta)}|j-G|^2-CR^2.
$$
This gives the desired result.
\cqfd

\begin{lem}
\label{newlem2}
Assume $j$ satisfies \eqref{equation1} and \eqref{equation2} with $\nu$  of uniform type and let $G$ be the function defined in Lemma \ref{lem1} --- for some   $\delta < \frac12 \inf\{|p-q|\mid p\neq q\in\Lambda\}$ --- and extended by $0$ on $\R^2\setminus \cup_{p\in \Lambda}B(p,\delta)$. Then 
$$
W(j)<\infty\;\Leftrightarrow\; \limsup_{R\to\infty} \dashint_{B_R}|j-G|^2<\infty
$$
where $\dashint_A$ denotes the average over $A$.
\end{lem}

\pr
The ``$\implies$" part of the assertion follows from Lemma \ref{lem1}. We prove the reverse implication. We denote by $g$ the result of applying Proposition~\ref{propprel} to $j$. 

Then from the properties of $\chi_R$ and using \eqref{error}, \eqref{lowbound}, 
$$W(j,\chi_R)\le \int_{B_R} g \chi_R + CR^2\le CR^2 + \int_{B_R\setminus \cup_{p\in \Lambda}B(p,\delta)} g+\sum_{p\in\Lambda\cap B_R}\int_{B(p,\delta)} g.$$ 
Then, as in the proof of Lemma~\ref{lem1}, 
$$\int_{B(p,\delta)} g = W(j,\indic_{B(p,\delta)}) = \frac12\int_{B(p,\delta)}|j-G|^2+O(1).$$ 
Using this and \eqref{outside} we find 
$$W(j,\chi_R)\le CR^2 + \frac12\int_{B_{R+\delta}}|j-G|^2.$$
This yields the desired result.
\cqfd

\pr[Proof of Proposition \ref{prop2}] We denote $\Lambda_i$  the lattice related to $j_i$ for $i=1,2$ and $\Lambda$ one  related to $j_1+j_2$. We write $\nu_i = \sum_{p\in \Lambda_i} \alpha_{i,p} \delta p$, $i=1,2$.   Then we choose 
$$\delta <\frac12 \min\(\inf\{|p-q|\mid p\neq q\in\Lambda_1\}, \inf\{|p-q|\mid p\neq q\in\Lambda_2\}, \inf\{|p-q|\mid p\neq q\in\Lambda\}\),$$
and let $G_i(x) = \alpha_{i,p}\frac{x-p}{|x-p|^2}$ if $x\in B(p,\delta)$ for $p\in\Lambda_i$, and $G_i = 0$ elsewhere. 

Then, under the assumptions of the proposition, there exists $C>0$ such that for any $R>0$ 
$$\int_{B_R} |j_1 - G_1|^2, \int_{B_R} |j_2 - G_2|^2 < C R^2.$$
Therefore 
$$\int_{B_R} |j_1+ j_2 - (G_1+G_2)|^2 < C R^2.$$
In view of the previous Lemma, Proposition \ref{prop2} is proved.
\cqfd

\begin{cor} 
\label{cor2}
We have, with $m=0$, 
$$
\tilde W(\Z) <+\infty,\;\tilde W( \N)<+\infty
$$
 \end{cor}
 \pr There exists $j_1\in \mathscr F_{\Z^2}$ and from Corollary~\ref{cor1} there exists $j_2\in\mathscr F_{\Z^2\setminus\Z}$ such that $W(j_1)$ and $W(j_2)$ are both finite with $m=1$.
 Then, by Proposition \ref{prop2} and since $-\div(j_1-j_2) = \sum_{p\in\Z} \delta_p$, and $\Z$  is  uniform, we have $W(j_1-j_2) < +\infty$ with $m=0$, hence $\tilde W(\Z)<+\infty$. The proof for $\N$ is identical.
 \cqfd

\begin{pro}
\label{prop4}
For $m=0$ we have
$$
 W( \Z)<+\infty,\; W( \N)=+\infty
$$
\end{pro}
\pr[The case of $\Z$.] We define $V_1(x):=-\log|\sin (\pi x)|$. Direct calculations lead to
 $$
 -\triangle V_1=2\pi \sum_{p\in \Z}\delta_p \mbox{ in
}\R^2
 $$
 and
 $$
 |\nabla V_1(x)|=\pi\frac{|\cos (\pi x)|}{|\sin (\pi x)|}.
 $$
 Both $V_1(x)$ and $ |\nabla V_1(x)|$ are $1$-periodic functions. Straightforward calculations yield 
 $$
 W(\nabla V_1)<+\infty.
 $$
 \cqfd
 
\pr[The case of $\N$.] We must prove that no  $\nabla U\in\mathscr P_\N$ is such that $W(\nabla U)<+\infty$. Our strategy is to construct $\nabla H_1\in\mathscr P_\N$ such that $W(\nabla H_1) = +\infty$, and such that $W(\nabla H_1,\chi_R) < CR^2\log^2R$. Then, if there existed $\nabla H_2\in\mathscr P_\N$  such that $W(\nabla H_2)<+\infty$, we would conclude that $W(\nabla (H_1-H_2),\chi_R)$ grows at most like $R^2\log^2 R$. Since $H_1-H_2$ is harmonic we conclude from a Liouville type theorem that  $\nabla (H_1-H_2)$ is constant, which contradicts $W(\nabla H_1) = +\infty$.

To construct $H_1$ we use the Weierstass construction for a holomorphic function in
the plane with  a simple zeroe at each $p\in\N$ to define 
\[
H(x):=\Pi_{k\in\N
}(1-\frac{x}{k})e^{\frac{x}{k}}.
\]
Then we let 
\[
H_1(x)=-\log |H(x)|.
\]
It is straightforward to check that the product in the definition of $H$ converges uniformly on any compact subset of $\C$ and that 
$$
-\triangle H_1 =2\pi \sum_{k\in \N}\delta_{k}\mbox{ in }\R^2
$$
and for all $x\in \C=\R^2$
\beq
\label{primaryresult1}
|H_1(x)|\le \sum_{k\in\N}\left|\log(1-\frac{x}{k})+\frac{x}{k}\right|
\eeq
and
\beq
\label{primaryresult2}
|\nabla H_1(x)|=\left| \sum_{k\in\N}\frac{x}{k(k-x)}\right|.
\eeq

Next, rather than proving $W(\nabla H_1,\chi_R) < CR^2\log^2R$, we prove the stronger, pointwise estimates:
\beq\label{supout} |\nabla H_1(x)|\le  C(\log(|x|+1) +1),\quad \text{outside $\cup_{k\in\N}B(k,\frac14)$},
\eeq
\beq\label{supin} \left|\nabla H_1(x)+\frac{1}{x-k}\right|\le  C(\log (|x|+1)+1),\quad \text{in $B(k,\frac14)$}.
\eeq
For \eqref{supout},  take  any $x\in\C\setminus\cup_{k\in\N}B(k,\frac14)$, it follows from (\ref{primaryresult2}) that 
\[
|\nabla H_1(x)|\le \sum_{1\le k\le [2|x|+1]}\(\left|\frac{1}{k-x}\right|+\left|\frac{1}{k}\right|\)+\sum_{k> [2|x|+1]}\left|\frac{x}{k(k-x)}\right|:=I+II,
\]
where $[\cdot]$ denotes the integer part of a real number. We have
\[
II\le \sum_{k> [2|x|+1]}\frac{|x|}{(k-|x|)^2}\le |x|\int_{|x|}^{+\infty} \frac{dt}{t^2} \le 1,
\]
\[
\sum_{1\le k\le [2|x|+1]}\frac{1}{k}\le1+ \int_{1}^{2|x|+1}  \frac{dt}{t} \le 2(\log(|x|+1)+1).
\]
On the other hand, 
\[
\sum_{1\le k\le [2|x|+1]}\left|\frac{1}{k-x}\right|\le\sum_{1\le k\le [2|x|+1]}\left|\frac{1}{{\cal Re}(k-x)}\right| \le 5+2\int_{1}^{2|x|+1}  \frac{dt}{t} \le  5(\log(|x|+1)+1).
\]
Therefore, for any $x\in\C\setminus\cup_{k\in\N}B(k,\frac14)$, we have $|\nabla H_1(x)|\le  8(\log(|x|+1)+1),$
and therefore \eqref{supout} holds.

Now we prove \eqref{supin}. Let $x\in B(k,\frac14)$ for some $k\in\N$.  As above
\[
\left|\nabla H_1(x)+\frac{({\cal Re}(x)-k, -{\cal Im}(x))}{|x-k|^2}\right|\le  8(\log (|x|+1)+1)+\frac1k\le 9(\log (|x|+1)+1),
\]
or equivalently, if we use the division of complex number,
$$\left|\nabla H_1(x)+\frac{1}{x-k}\right|\le  8(\log (|x|+1)+1)+\frac1k\le 9(\log (|x|+1)+1),$$
since $x\in\C\setminus\cup_{i\neq k\in\N}B(i,\frac14)$. This proves \eqref{supin}

We now turn to the proof that $W(\nabla H_1) = +\infty$. This is done by computing a lower bound for $|\nabla H_1(x)|$. More precisely we prove that  or any $\e>0$, there exists some positive constant $C_1$  depending on $\e$ such that 
\beq
\label{basicest}
|\nabla H_1(x)|\ge ( \log (|x|+1)-C_1),\quad \text{if $|{\cal Im}(x)|\ge \e |x|+1$.}
\eeq
For this purpose  we consider the meromorphic function 
\[
f(x):=\sum_{k\in\N}\frac{x}{k(k-x)}.
\]
If $|{\cal Im}(x)|\ge \e |x|+1$, then $x\in\C\setminus\cup_{k\in\N}B(k,\frac14)$. Thus 
\[
\left|f(x)-\sum_{1\le k\le [2|x|+1]}\(\frac{1}{k-x}-\frac{1}{k}\)\right| \le II \le 1,
\]
so that
\[
\begin{array}{llll}
\ds\left|f(x)+\sum_{1\le k\le [2|x|+1]}\frac{1}{k}\right| &\le&\ds 1+\sum_{1\le k\le [2|x|+1]}\left|\frac{1}{k-x}\right|\\
&\le&\ds  1+\sum_{1\le k\le [2|x|+1]}\left|\frac{1}{{\cal Im}(x)}\right|\\
&\le&\ds  1+\sum_{1\le k\le [2|x|+1]}\frac{1}{|{\cal Im}(x)|}\\
&\le&\ds  1+\frac{2|x|+1}{|{\cal Im}(x)| }\\
&\le &1+2/\e.
\end{array}
\]
On the other hand, we have
$$
\sum_{1\le k\le [2|x|+1]}\frac{1}{k}\ge \log (|x|+1),
$$
hence (\ref{basicest}) follows. We claim that this implies that $W(\nabla H_1)=+\infty$.

To see this, we need to bound from below the integral of $\chi_R|\nabla H_1|^2$. We define $g$ by applying Proposition~\ref{propprel} to $\nabla H_1$ with $\delta = 1/4$. Then we deduce from \eqref{lowbound}, \eqref{outside} and the fact that $\chi_R = 1$ on $B_{R-1}$ that 
$$\int \chi_R |\nabla H_1|^2\ge \int_{B_{R-1}} |\nabla H_1|^2 - CR.$$
Then, integrating \eqref{basicest} on $\{x\in B_{R-1}\mid |{\cal Im}(x)|\ge \e |x|+1\}$ proves that $W(\nabla H_1)=+\infty$.

We may now argue by  contradiction to prove the proposition. Assume that there exists $H_2\in\mathscr P_\N$ such that $W(\nabla H_2)<+\infty$. Then $\bar H = H_2 - H_1$ is a harmonic function over $\R^2$. For $i=1,2$ we define $g_i$ by applying Proposition~\ref{propprel} to $\nabla H_i$ with $\delta = 1/4$. Then 
$$CR^2\ge W(\nabla H_2, \chi_R)- W(\nabla H_1, \chi_R) \ge \int \chi_R (g_2 - g_1) - CR\ge \int_{B_{R-1}} (g_2 - g_1) - CR.$$
Then, letting $G(x) = (x-k)/|x-k|^2$ in $B(k,1/4)$ for every $k$ and $G = 0$ outside $\cup_k B(k,1/4)$ we have, as in \eqref{eq2.1.3}, for every $k$
$$ \int_{B(k,1/4)} g_i = \int_{B(k,1/4)} \frac12 |\nabla H_i - G|^2 + C_0,$$
where $C_0 = -\pi\log 4 $. Together with \eqref{outside}, this implies that 
$$\int_{B_{R-1}} (g_2 - g_1) \ge \frac12\int_{B_{R-1}\setminus\cup_k B(k,1/4)} \(|\nabla H_2|^2 - |\nabla H_1|^2\) - \frac12 \sum_{k=0}^{[R]}\int_{B(k,1/4)} \frac12 |\nabla H_1 - G|^2  - CR.$$
Using \eqref{supin} we have 
$$\int_{B(k,1/4)} \frac12 |\nabla H_1 - G|^2\le C\(\log(k+1)+1\)^2,$$ 
so that 
$$ CR^2\ge \frac12\int_{B_{R-1}\setminus\cup_k B(k,1/4)} \(|\nabla H_2|^2 - |\nabla H_1|^2\) - CR\log^2R.$$
Then, writing 
$$|\nabla H_2|^2 - |\nabla H_1|^2 = |\nabla \bar H|^2 + 2\nabla \bar H\cdot\nabla H_1,$$
we find using \eqref{supout} that on $B_{R-1}\setminus\cup_k B(k,1/4)$
$$|\nabla H_2|^2 - |\nabla H_1|^2 \ge |\nabla \bar H|^2 - C\log R|\nabla \bar H|,$$
and thus, letting $A_R = B_{R-1}\setminus\cup_k B(k,1/4)$, 
$$ CR^2\ge \frac12\int_{A_R} \(|\nabla \bar H|^2 - C\log R|\nabla \bar H|\) - CR\log^2R,$$
from which we easily deduce
$$\int_{A_R}|\nabla \bar H|^2\le C R^2\log^2 R.$$
It follows by a mean value argument that there exists $t\in [R/2, R-1]$ such that 
$$\int_{\partial B_t}|\nabla \bar H|^2\le C R\log^2 R,$$
and since $\bar H$ is harmonic, for any $x\in B_{R/4}$ we have 
$$|\nabla^2 \bar H(x)|\le \frac1{R^2} \int_{\partial B_t}|\nabla \bar H|\le C \frac1{R^2} \sqrt R R\log R.$$
Fixing $x$ and letting $R\to\infty$, we find $\nabla^2 \bar H(x)=0.$ Therefore $\nabla \bar H$ is a constant, which is clearly not possible since $W(\nabla H_1)=+\infty$ while $W(\nabla H_1+\nabla \bar H)<+\infty$.
\cqfd

We summarize the content of this section in the following 
\begin{thm}
\label{thm1}
We have 
\begin{eqnarray}
\label{energy1}
\tilde W(\Z)<+\infty,\;\tilde W(\N)<+\infty,\; \tilde W(\Z^2\setminus \Z)<+\infty,\; \tilde W(\Z^2\setminus \N)<+\infty\\
\label{energy2}
 W(\Z)<+\infty,\;W(\Z^2)<+\infty,\;  W(\Z^2\setminus \Z)<+\infty\\
\label{energy3}
 W(\N)=+\infty,\;  W(\Z^2\setminus \N)=+\infty 
\end{eqnarray}
\end{thm}

\pr
The result comes from Corollary \ref{cor1}, Corollary \ref{cor2}, Proposition \ref{prop2} and Proposition \ref{prop4}.
\cqfd

\section{Sufficient conditions for finite  renormalized energy}
\begin{thm}
\label{thm2}
Given a discrete lattice $\Lambda$,  assume there exists  $m\ge 0$ and 
 $\e\in (0,1)$, $C>0$   such that for any $x\in\R^2$ and for  $R>1$, we have
\beq\label{eps}
\left| \sharp\(B(x,R)\cap \Lambda\)-m\pi R^2\right|\le CR^{1-\e}
\eeq
 and
\beq\label{sep}
  \inf_{x,y\in\Lambda, x\neq y}|x-y|>0
\eeq
Then $W(\Lambda)<+\infty$.
\end{thm}
\begin{rem}
For a Bravais lattice, the assumptions in the above theorem are satisfied. It was proved by Landau (1915) --- see \cite{reseaux} for a more general statement --- that the first assumption  holds with $\e = 1/3$, see \cite{gotze} for references on more recent developments.
\end{rem}

We recall a technical lemma.\\
\begin{lem}(Theorem 8.17 in \cite{GT})
\label{lem3.1}
Assume $q>2$ and $p>1$ and $v$ is a solution of the following equation
$$
-\triangle u = g+\sum_i\partial_i f_i
$$
there exists some
constant $C$ such that
$$
\|u\|_{L^\infty(B(0,R))}\le C(R^{-\frac2p}\|u\|_{L^p(B(0,2R))}+R^{1-\frac2q}\|f\|_{L^q(B(0,2R))}+R^{2-\frac4q}\|g\|_{L^{q/2}(B(0,2R))})
$$
\end{lem}

\pr[Proof of Theorem \ref{thm2}.] Assume $\Lambda$ satisfies \eqref{eps} and \eqref{sep}. The proof consists in constructing $j\in\Fl$ such that $W(j)<+\infty$, which is done by successive approximations constructing a first some $U^1$, then a correction $U^2$ to $U^1$, then a correction $U^3$ to $U^1+U^2$, etc... In this construction, the $U^k$'s are functions, and  the sum of their gradients will converge to $j$.

Let $R_n=2^{n-1}$. For all $p\in \Lambda$, we let $U^1_p$ be the solution to 
\[
\left\{
\begin{array}{lllll}
-\triangle U^1_p(y) &=&\ds2\pi\left(\delta_{p}(y)-\frac{\indic_{B(p,R_1)} (y)}{\pi {R_1}^2}\right)&\mbox{ in
}B(p,R_1) \\
 U^1_p(y)=\ds\frac{\p U^1_p}{\p \nu}(y)&=&0& \mbox{ on }\p B(p,R_1) 
\end{array}
\right.
\]
where  $\indic_{B(x,r)}$ is the indicator function of the ball $B(x,r)$. The existence of a solution with Neumann boundary conditions follows from the fact that $\delta_{p}-\frac{\indic_{B(p,R_1)}}{\pi {R_1}^2}$ has zero integral, and the radial symmetry of the solution implies $U^1_p$ is constant on the boundary, and the constant can be taken equal to zero. 
In fact, extending $U^1_p$ by zero outside $B(p,R_1) $, we get a solution of  
$$-\triangle U^1_p(y) =\ds2\pi\left(\delta_{p}(y)-\frac{\indic_{B(p,R_1)} (y)}{\pi R_1^2}\right)$$
in $\R^2$, which is supported in $B(p,R_1)$. 

We let 
\[
U^1(y):=\sum_{p\in \Lambda} U^1_p(y).
\]
This sum is well defined since, $\Lambda$ being discrete,  it is locally finite. Moreover $U^1$ solves
\beq\label{equn}
-\triangle U^1(y) =2\pi\(\sum_{p\in
\Lambda}\delta_{p}-n_1(y)\),\quad\text{where}\quad
n_1(y):=\frac{\sharp\(\Lambda \cap B(y,R_1)\)}{\pi {R_1}^2}
\eeq
Then we proceed by induction. For any $k\ge 2$ we let $U^k$ be the solution to 
\beq\label{equm}
\left\{
\begin{array}{lllll}
-\triangle U^k_p(y) &=&\ds2\pi\left(\frac{\indic_{B(p,R_{k-1})}(y)}{\pi {R_{k-1}}^2}-  \frac{ \indic_{B(p,R_k)}(y)}{\pi {R_k}^2}  \right)&\mbox{ in
}B(p,R_k)  \\
U^k_p(y)  =\frac{\p U^k_p}{\p \nu}(y)&=&0& \mbox{ on }\p B(p,R_k), 
\end{array}
\right.
\eeq
and we let $ U^k_p = 0$  outside the $B(p,R_k)$. We let $U^k(y):=\sum_{p\in\Lambda} U^k_p(y)$, so that
$$ -\triangle U^k(y) =2\pi\(n_{k-1}(y)-n_k(y)\),$$
where, for any $k\in\N$, 
\[
n_k(y):=\frac{\sharp\(\Lambda \cap B(y,R_k)\)}{\pi R_k^2}.
\]
Now we study the convergence of  $\sum_{k=1}^\infty \nabla U^k$.

First we note that there is an explicit formula for $U^k_p$. For any $k\ge 2$ we have
\[
U^k_p(y)=V\(\frac{y-p}{R_k}\),\quad \text{where}\quad V(y):=\left\{
\begin{array}{llll}
-\frac{3|y|^2}2+\ln 2&\mbox{if }|y|\le \frac 12\\
\frac{|y|^2}2-\ln|y|-\frac12&\mbox{if }\frac12<|y|\le 1\\
0&\mbox{if }|y|\ge 1,
\end{array}
\right.
\]
from which it follows,  since $\|\nabla U^k_p\|_\infty\le \frac C{R_k}$ and the sum defining $U^k$ has at most  $CR_k^2$ non zero terms, that 
\beq\label{estgrad}\|\nabla U^k\|_\infty\le CR_k.
\eeq

Second we estimate $\| U^k\|_\infty$. We claim that 
\beq\label{estU} \text{$\forall k\ge 2$, $\exists C_k\in\R$ such that $\| U^k(y)-C_k\|_\infty=O(R_k^{1-\e})$.}
\eeq
Indeed,  from \eqref{eps},  
\beq\label{nk}
\|n_k - m\|_\infty \le C R_k^{-1-\e}.
\eeq
On the other hand, letting $n_y(r):=\sharp\(B(y,r)\cap\Lambda\)$, we have for any $y\not\in\Lambda$ 
$$
U^k(y)=\sum_{p\in B(y,R_k)\cap \Lambda}V\(\frac{|p-y|}{R_k}\)=\int_{0}^{R_k}V\(\frac{t}{R_k}\)n'_y(t)dt=-\int_{0}^{R_k}\frac1R_kV'\(\frac{t}{R_k}\)n_y(t)dt.
$$
But, using \eqref{eps},  we have $n_y(t)= m\pi t^2+O(t^{1-\e})$, hence
$$
U^k(y)=-m\pi\int_{0}^{R_k}\frac1R_kV'\(\frac{t}{R_k}\)t^2 dt+O(R_k^{1-\e}).
$$
The first term is independent of $y$, we call it $C_k$. This   proves \eqref{estU}. 

Finally, we note that,  from \eqref{nk}, it holds that 
\beq\label{estlap}\|\triangle U^k\|_\infty=O(R_k^{-1-\e}).\eeq

Now, we claim that \eqref{estU} and \eqref{estlap} imply that 
\beq\label{est}\|\nabla U^k\|_\infty=O(R_k^{-\e})\eeq
To see this we use the elliptic estimate of Lemma \ref{lem3.1}.  For all $y\in\R^2$ we have
\begin{multline}\label{estl2}
\ds \int_{B(y,R_k)} |\nabla U^k|^2=\int_{B(y,R_k)} |\nabla (U^k-C_k)|^2 
\\ =\ds -\int_{B(y,R_k)} \triangle U^k (U^k-C_k)+\int_{\p B(y,R_k)} \frac{\p U^k}{\p \nu} (U^k-C_k)\le  C {R_k}^2\(R_k^{-2\e}+\|\nabla U^k\|_\infty {R_k}^{-\e}\)
\end{multline}

Now we apply Lemma \ref{lem3.1}. For $i=1,2$ we have 
$$\triangle \(\partial_i U^k\) = -2\pi \partial_i\(n_{k-1} - n_k\),$$
therefore for any $q>2$ and $p>1$, 
$$\|\partial_i U^k\|_{L^\infty(B_{R_k/2})} \le C\({R_k}^{-\frac2p} \|\partial_i U^k\|_{L^p(B_{R_k})} + {R_k}^{1-\frac2q} \|n_{k-1} - n_k\|_{L^q(B_{R_k})}\).$$
Then,  taking $p=2$ and noting that  \eqref{nk} implies  $\|n_{k-1} - n_k\|_q\le C {R_k}^{\frac2q-(1+\e)}$, we find using \eqref{estl2} that 
$$\|\partial_i U^k\|_{L^\infty(B_{R_k/2})} \le C\({R_k}^{-2\e} + {R_k}^{-\e} \|\nabla U^k\|_{L^\infty(B_{R_k})}\)^{\frac12}  + C {R_k}^{-\e}.$$
This proves \eqref{est}. 

Now \eqref{est} implies that the sum $\sum_{k\ge 2} \nabla U^k$ converges, and if we let $j=\nabla U_1 +\sum_{k\ge 2} \nabla U^k$, then $-\div j = 2\pi(\sum_{p\in\Lambda} \delta_p -m)$, using \eqref{equn}, \eqref{equm} and \eqref{nk}. Moreover $j$ is a gradient since it is a sum of gradients, thus $j\in\Pl$. 

To conclude, it is easy to check, using the assumption $\inf_{x,y\in\Lambda, x\neq y}|x-y|>0$, that  $W(\nabla U^1, \chi_R)\le CR^2$ for all $R>1$, and to deduce using \eqref{est} that $W(j, \chi_R)\le CR^2$.
\cqfd

For $\tilde W$ the hypothesis of Theorem \ref{thm2} can be relaxed  somewhat.\\

\begin{thm2bis}
\label{thm2bis}
Assume there exists some non-negative number $m\ge 0$ and some positive numbers 
 $\e\in (0,1)$, $C>0$  and a increasing sequence $\{R_n\}$ tending to $+\infty$ such that for any $x\in\R^2$ and for any $n\in\N$, we have
 \[
\left|\sharp\(B(x,R_n)\cap \Lambda\)-m\pi R_n^2\right|\le CR_n^{1-\e},
 \]
 and such that 
 \[
\sum_n R_n^{-\e}<+\infty
 \]
 and 
  \[
  \inf_{x,y\in\Lambda, x\neq y}|x-y|>0.
  \]
Then $\tilde W(\Lambda)<+\infty$.
\end{thm2bis}

We will use the following simple  estimate.\\
\begin{lem}
\label{lem2}
Let $u$ be a solution of the following problem
\[
\left\{
\begin{array}{lllll}
-\triangle u &=&f&\mbox{ in
}\Omega \\
\frac{\partial u }{\partial \nu}&=&0& \mbox{ on }\partial\Omega
\end{array}
\right.
\]
Then
\[
\int_{\Omega}|\nabla u|^2\le C|\Omega|^2\|f\|_\infty^2
\]
where $C$ is a constant independent of $\O$.
\end{lem}
\pr We have
$$
\int_{\Omega}|\nabla u|^2=-\int_{\Omega}u\triangle
u=\int_{\Omega}f u \le \|u\|_1\|f\|_\infty\le
\sqrt{|\Omega|}\|u\|_2\|f\|_\infty
$$
Without loss of generality, we assume $\int u=0$. By Poincar\'e
inequality,
$$
\|u\|_2\le C\sqrt{|\Omega|}\|\nabla u\|_2.
$$
Finally, the desired result follows.
\cqfd

\pr[ Proof of Theorem $2'$] Let 
$$\mul = \sum_{p\in\Lambda}\delta_p,\quad I_k = \frac{\indic_{B_{R_k}}}{|B_{R_k}|},$$
for any integer $k>0$, where $\indic_{B_{R_k}}$ is the indicator function of the ball $B(0,R_k)$. 

At the first step, for all $x\in\R^2$ we let $U^1_x$ be the solution to 
\[
\left\{
\begin{array}{lllll}
-\triangle U^1_x(y) &=&\ds2\pi\left(\mul(y)- \mul * I_1(x)\right)\indic_{B_{R_1}} (x-y)&\mbox{ in
}B(x,R_1+1) \\
 \frac{\p U^1_x}{\p \nu}(y)&=&0& \mbox{ on }\p B(x,R_1+1).
\end{array}
\right.
\]
This equation has a solution which is unique up to an additive constant since
$$\int \left(\mul(y)- \mul * I_1(x)\right)\indic_{B_{R_1}} (x-y)\,dy = \mul * (\pi {R_1}^2 I_1)(x) - \pi {R_1}^2\mul * I_1(x)=0.$$
We extend $\nabla U^1$ by zero outside $B(x,R_1+1)$ and let 
\[
j^1(y):=\frac1{\pi{R_1}^2}\int_{\R^2}\nabla U^1_x(y)dx, 
\]
so that 
\[
-\div(j^1) =2\pi\(\sum_{p\in\Lambda}\delta_{p}-m_1(y)\), \quad\text{where}\quad
m_1 =\mul * I_1 * I_1.
\]

Then we define $j^k$ by induction. For $x\in\R^2$ we let $U^k_x$ be the solution to 
\[
\left\{
\begin{array}{lllll}
-\triangle U^k_x(y) &=&\ds 2\pi\left(m_{k-1}(y)- m_{k-1} * I_k(x)\right)\indic_{B_{R_k}} (x-y)&\mbox{ in
}B(x,R_k+1)  \\
  \frac{\p U^k_x}{\p \nu}(y)&=&0& \mbox{ on }\p B(x,R_k+1), 
\end{array}
\right.
\]
and extend $\nabla U^k_x$ by  $0$ outside the ball $B(x,R_k)$. Then we let 
\[
j^k(y):=\frac{1}{\pi R_k^2}\int_{\R^2}\nabla U^k_x(y)dx
\]
so that 
\[
-\div(j^k)(y) =2\pi(m_{k-1}(y)-m_k(y)),\quad\text{where}\quad
m_k=m_{k-1}*I_k*I_k.
\]
We claim that 
\beq\label{mk}
m_k(y)=m+O(R_k^{-1-\e}).
\eeq
To see this, it suffices to note that from the commutativity of the convolution we have 
$$ m_k=(\mul * I_k) * (I_k*I_{k-1} * I_{k-1} *\dots * I_1*I_1).$$
Then from our first assumption $|\mul * I_k - m| \le C {R_k}^{-(1+\e)}$, which implies \eqref{mk} since every $I_k$ is a positive function with integral $1$, and thus convoluting a function with it does not increase the $L^\infty$ norm. 

It follows from  Lemmas \ref{lem3.1} and \ref{lem2} that  for all $k\ge 2$ and $x\in\R^2$
\[
\|\nabla U^k_x\|_{L^\infty(\R^2)}\le C {R_{k-1}}^{-\e},
\]
which yields for all $k\ge 2$
\[
\| j^k\|_{L^\infty(\R^2)}\le C {R_{k-1}}^{-\e}.
\]
Therefore $\sum_{k\ge 2}\| j^k\|_\infty\le +\infty$ and we can define $j:= \sum_{k\ge 1}j^k$. The vector field $j$ solves
\[
-\div(j) =2\pi\(\sum_{p\in \Lambda}\delta_{p}-m\)\qquad\mbox{in }\R^2.
\]
Now it suffices to prove that $W(j)<+\infty$. This is  clearly a consequence of the fact that $W(j_1)<+\infty$ and the fact that 
 $\sum_{k\ge 2}\| j^k\|_\infty\le +\infty$. On the other hand, $W(j_1)<+\infty$ is proved as follows: For any $p\in\Lambda$, and any $x\in B(p,R_1)$ we have  $\|U^1_x(y)-log|y-p|\|<C$ in  $C^1(B(p,\delta))$ with $C,\delta>0$ independent of  $p$, $x$ and $y$, because of the equation satisfied by $U^1$ and the uniform spacing of the points in $\Lambda$. Also, if $x\notin B(p,R_1)$, then $\|U^1_x(y)\|<C$ in  $C^1(B(p,\delta))$. 
 
Then, since $j^1 = \int\nabla U^1_x/\pi {R_1}^2$, we have $|j^1(y)-log|y-p||<C$ in  $B(p,\delta)$ for any $p\in\Lambda$ and $|j^1|< C$ outside $\cup_{p\in\Lambda}B(p,\delta)$. This implies that $W(j_1)<+\infty.$
\cqfd

\begin{pro}\label{opt}
The conditions in Theorem \ref{thm2} are optimal in some sense. More precisely, for any $m\ge 0$ and any $\e>0$ there exists $\Lambda$ such that  $\tilde W(\Lambda) = +\infty$ and for any $x\in\R^2$ and any $R>1$
$$\left| \sharp\(B(x,R)\cap \Lambda\)-m\pi R^2\right|\le CR^{1+\e}.$$
\end{pro}

\pr The counter-example is as follows, assuming without loss of generality that $\e<1/2$:  For all $k\in \N$, on the circle $\partial B(0,4k)$, we distribute uniformly $[32\pi m k + k^{\e}]$ points,  where $[x]$ is the integer part of $x$. This is clearly possible maintaining at the same time a distance greater than $\min(1/5m,1)$ (if $k$ is large enough) between the points, since $k^\e \ll k$ as $k\to +\infty$.

Then we have as $k\to +\infty$
$$\sharp\(\Lambda\cap B(0,4k)\) -m\pi (4k)^2\simeq \sum_{i=1}^{k-1} [i^{\e}] \simeq \frac{k^{1+\e}}{1+\e},$$
thus for any $j$ such that 
$$
-\div(j)=2\pi\(\sum_{p\in \Lambda} \delta_p -m\),
$$
and for any $R\in (4k+1, 4k+3)$, we have
$$
\int_{\partial B(0,R)} j\cdot \nu =2\pi \(\sharp\(\Lambda\cap B(0,R)\)-m\pi R^2\)\simeq \frac{2\pi}{1+\e}k^{1+\e}.
$$
Thus there exist $k_0>0$, $c_0>0$ such that if $k>k_0$ and for any $R\in (4k+1, 4k+3)$, we have
\beq\label{ests1}
\frac12\int_{\partial B(0,R)}|j|^2\ge\frac1{4\pi R} \left( \int_{\partial B(0,R)} j\cdot \nu  \right)^2\ge c_0 k^{1+2\e}
\eeq
Now we construct $g$ using proposition~\ref{propprel} with $\delta < \frac12 \inf_{p\neq q\in\Lambda}|p-q| $ and $\delta<1$. For functions $\{\chi_R\}_R$ satisfying \eqref{chir}, we have for any $k\in\N$ and since the support of $\chi_{4k+2}$ does not intersect $\cup_{p\in\Lambda} B(p,\delta)$ that 
$$W(j,\chi_{4k+2}) = \int g\chi_{4k+2} = \int_{\cup_{p\in\Lambda} B(p,\delta)} g \chi_{4k+2} + \int_{\R^2\setminus\cup_{p\in\Lambda} B(p,\delta)} g \chi_{4k+2}$$
and therefore,  since $g = \frac12 |j|^2$ outside $\cup_{p\in\Lambda} B(p,\delta)$ and $g\ge -C$, we obtain
$$
W(j,\chi_{4k+2})\ge \sum_{i\le k-1}\int_{B(0,4i+3)\setminus B(0,4i+1)}|j|^2-CR^2\ge CR^{2+2\e},
$$
where we used \eqref{ests1} for the last inequality. Therefore $W(j)=+\infty.$
\cqfd

\section{Critical case}

In view of Theorem~\ref{thm2} and Proposition~\ref{opt}, the critical discrepancy between $\sum_{p\in\Lambda}\delta_p$ and the uniform measure $m\,dx$ is when $\left| \sharp\(B(x,R)\cap \Lambda\)-m\pi R^2\right|= O(R).$ This includes the cases $\Lambda = \Z$ or $\N$. As shown by Theorem~\ref{thm1}, we cannot expect $W(\Lambda)$ to be finite under such an assumption. However we have the following result for $\tilde W$. 

\begin{thm}
\label{critical}
Let $A\subset \Z^2$ and $\Lambda:= \Z^2\setminus A$. Assume there exists some constant $C>0$ such that for all $x\in \R^2$ and for all  $R>1$  we have
$$
\sharp\(A\cap B(x,R)\)\le CR.
$$
Then 
$$
\tilde W(\Lambda)<+\infty
$$
\end{thm}

This result is a direct consequence of  Proposition  \ref{pro-move} and the following:

\begin{pro}
\label{procritical} Let $A\subset \Z^2$. Then the following properties are equivalent.

\begin{description}
\item[Property I.] There exists some constant $C>0$ such that for all $x\in \R^2$ and for all  $R>1$  we have
\beq\label{disc}\sharp\(A\cap B(x,R)\)\le CR\eeq
\item[Property II.]
There exists a bijective map $\Phi: \Lambda\to \Z^2$ satisfying
\beq\label{phi}
  \sup_{p\in \Lambda}  |\Phi(p)-p|<\infty
\eeq
 \end{description}
\end{pro} 

The fact that the second property implies the first one is not difficult. First note that \eqref{phi} is equivalent to the same property for $\Phi^{-1}$, and that Property~I is equivalent to the same property with squares $K_R$ of sidelength $R$ replacing the balls of radius $R$.

Now assume $\sharp\(A\cap K_R\)> CR,$ then $\Phi^{-1}(K_R\cap\Z^2)$ is included in $\Z^2\setminus A$ and thus contains at least $CR$ points which do not belong to $K_R$. Therefore, as $C\to +\infty$, the maximal distance between an element $p$ of $\Phi^{-1}(K_R\cap\Z^2)$ and $K_R$ tends to $+\infty$. This proves that II $\implies$ I.

The proof of the converse is less obvious. It is essentially an application of the max-flow/min-cut duality, with arguments similar in spirit to those found in \cite{strang}. 

We let ${\cal G}$ be a graph for which the set of vertices is $\Z^2$ and the set of edges is 
 \[
  {\cal A}:=\{ (p,q)\quad|\quad p,q\in \Z^2, \|p-q\|=1\}
 \]
 where $\|\cdot\|$ is the Euclidean norm. Given an integer $N\in\N$, we define some function
 \[
 \begin{array}{lllll}
 \mu_N:&\Z^2&\to&\R_+\\
 &p&\mapsto&N^2-\sharp(\Lambda\cap K^N_p)
 \end{array}
 \]
where $K^N_p:= [kN, (k+1)N)\times[lN, (l+1)N)$ for $p=(k,l)$. Since $\Lambda=\Z^2\setminus A$, $\mu_N$ is indeed  non-negative  and $ \mu_N(p)$ is equal to $\sharp(A\cap K^N_p)$, i.e. the deficit of the points of $\Lambda$ in $ K^N_p$.

We introduce the following notions.

\begin{itemize}
\item A {\it flow}, or {\it 1-form} is a map $\varphi: {\cal A}\to\R$ such that for any edge  $(p,q)$ one has $\varphi(p,q)=-\varphi(q,p)$.
\item Given a flow $\varphi$, its divergence $\div(\varphi)$ is the function $\div(\varphi): \Z^2\to \R$ such that for any $p\in \Z^2$ one has
  \[
 \div(\varphi)(p):=\sum_{(p,q)\in {\cal A}} \varphi (p,q)
  \]
\item Given a function $f:\Z^2\to\R$, its {\it gradient} $\nabla f$ is the 1-form $\nabla f(p,q)=f(q)-f(p)$.
\item Given a subset $A$ of $\Z^2$, its {\it boundary} $\partial A$ is defined by
 \[
 \partial A:=\{(p,q)\in \cal A\mid p\in A, q\in \Z^2\setminus A\}.
 \]
 \item Given a subset $A$ of $\Z^2$, its {\it perimeter} is  $\per(A):=\sharp (\partial A)$
 \item A {\it curve} connecting $p$ and $q$  is a subset of ${\cal A}$ of the form $\{(p_0,p_1),(p_1,p_2),\cdots, (p_{n-1},p_n)\}$ with $p_0=p$ and $p_n=q$.  A {\it loop}  or {\it cycle} is curve such that     $p_n=p_0$. A graph is {\it connected} if any two points can be connected by a curve.
 \item Given a function $f:\Z^2\to\R$ and $B\subset \Z^2$, its {\it integral}  on $B$ is defined by 
\[
\int_{B}f:= \sum_{p\in B} f(p).
 \] 
 We denote also $f(B)=\int_{B}f$.
 \item Given a 1-form $\varphi$ and a curve $\gamma$, the {\it integral} of $\varphi$ on $\gamma$ is defined by 
\[
\int_{\gamma}\varphi:= \sum_{a\in \gamma} \varphi(a)
 \] 
 \item Given two 1-forms $\varphi$ and $\phi$, their {\it inner  product} is
\[
\langle \varphi, \phi\rangle:=\frac12 \sum_{a\in {\cal A}}\varphi(a)\phi(a)
 \] 
  \item Given a  1-form $\varphi$ and a subset $S\subset {\cal A}$, the {\it total variation} of $\varphi$ with respect to $S$ is defined by 
\[
[ \varphi,S]:=\frac12 \sum_{a\in S}|\varphi(a)|
 \] 
 When $S={\cal A}$, we  simply write $[\varphi]$.
\end{itemize}

We have the following  classical results.

\begin{lem}
\label{Poincare}
(Poincar\'e Lemma) Given a 1-form $\varphi$, if one has $\int_\gamma\varphi=0$ for any loop $\gamma$, then there exists a function $f$ satisfying
\[
\varphi=\nabla f
\]
\end{lem}
\pr One fixes some point $p\in \Z^2$ and for any $q\in\Z^2$ one defines $f(q):=\int_{\mathscr C}\varphi$ where $\mathscr C$ is any curve connecting $p$ and $q$. From the hypothesis, this definition is independent of the particular curve  chosen, and it is easy to check that  $\varphi=\nabla f$.\cqfd

\begin{lem}
\label{Stokes}
(Stokes' formula) Let $\varphi$ be  a 1-form with compact support and $f$ be a function with  compact support. Then one has
\[
\langle \varphi,\nabla f\rangle =-\int_{\Z^2} f\,\div(\varphi)
\]
\end{lem}
\pr
We write $\varphi$ as linear combination  of elementary 1-forms
\[
\alpha_{(p,q)}:=\delta_{(\{p,q)\}}-\delta_{\{(q,p)\}},
\]
and note that 
\[
\langle \alpha_{(p,q)},\nabla f\rangle=f(q)-f(p) =-\int_{\Z^2} f\div(\alpha_{(p,q)}),
\]
since $\div(\alpha_{(p,q)})=\delta_{\{p\}}-\delta_{\{q\}}$.
\cqfd

\begin{lem}
\label{Coarea}
(Coarea formula) Let  $f:\Z^2\to \R_+$ be a function with the compact support. Then one has
\[
[\nabla f]=\int_0^\infty \per(\{f>t\}) dt
\]
\end{lem}
\pr
We note that 
\[
[\nabla f, \{(p,q),(q,p)\}]=|f(q)-f(p)|
\]
and 
\[
\partial \{f>t\}\cap  \{(p,q),(q,p)\}=
\left\{
\begin{array}{lllll}
(p,q)&\mbox{if } f(p)>t \mbox{ and }  f(q)\le t \\
(q,p)&\mbox{if } f(q)>t \mbox{ and }  f(p)\le t \\
\emptyset &\mbox{ otherwise, }
\end{array}
\right.
\]
which implies that 
\[
\sharp(\partial \{f>t\}\cap  \{(p,q),(q,p)\})=
\left\{
\begin{array}{lllll}
1 &\mbox{ if } f( p)>t\ge f(q)\mbox{ or } f( q)>t\ge f( p)\\
0 &\mbox{ otherwise }
\end{array}
\right.
\]
Therefore, we get
\[
\int_0^\infty  \sharp(\partial \{f>t\}\cap  \{(p,q),(q,p)\}) dt=|f(q)-f(p)|.
\]
Summing with respect to all couples of edges $\{(p,q),(q,p)\}$ proves the result.\cqfd

We may now set up the duality argument. For any given $1$-form $\varphi$ we let 
\[
\|\varphi\|_\infty=\sup_{(p,q)\in\cal A} \varphi(p,q) = \sup\{\langle \phi,\varphi\rangle\mid \text{$\phi$ is compactly supported, $[\phi]\le 1$}\}, 
\]
and  define, 
\[
\alpha:=\min_{-\div(\varphi)=\mu_N} \max\{\langle \phi,\varphi\rangle\mid \phi\in C_0, [\phi]\le 1\},
\]
where $C_0$ is the set of compactly supported $1$-forms. 
\begin{lem}
\label{express}
One has
\[
\alpha=\max_{\nabla f\in C_0, [\nabla f]\le 1}\int_0^{+\infty}\(\int_{\{f>t\}}\mu_N -\int_{\{f<-t\}}\mu_N \) dt
\]
\end{lem}
\pr
By  convex duality, we obtain
\beq\label{maxmin}
\alpha= \max_{\{\phi\in C_0| [\phi]\le 1\}}\min_{-\div(\varphi)=\mu_N} \langle \phi,\varphi\rangle.
\eeq
Then given  $\phi\in C_0$, we assume there exists a loop $\gamma$ such that
\[
\int_{\gamma} \phi\neq 0.
\]
We may then define $\varphi_t$ for any $t\in\R$ by
\[
\varphi_t(a):=\left\{
\begin{array}{lll}
t&\mbox{ if }a\in \gamma\\
0&\mbox{ otherwise }
\end{array}
\right.
\]
Since $\gamma$ is a loop, $\varphi_t$ has compact support and $\div(\varphi_t)=0$. Moreover, since $\int_{\gamma} \phi\neq 0$, 
\[
\min_{t\in \R} \langle \phi,(\varphi+\varphi_t)\rangle=-\infty
\]
which implies
\[
\min_{-\div(\varphi)=\mu_N} \langle \phi,\varphi\rangle=-\infty.
\]
As a consequence, the maximum in \eqref{maxmin} can be restricted to those $\phi$'s for which the   integral on any loop is zero, i.e. to gradients, in view of Lemma~\ref{Poincare}. Therefore 
\beq\label{maxmingrad}
\alpha= \max_{\{\nabla f\in C_0| [\nabla f]\le 1\}}\min_{-\div(\varphi)=\mu_N} \langle \nabla f,\varphi\rangle
\eeq

Now, from Lemma~\ref{Stokes}, we have
\[
\alpha= \max_{\{\nabla f\in C_0| [\nabla f]\le 1\}}\min_{-\div(\varphi)=\mu_N} \int -\div(\varphi)f=\max_{\{\nabla f\in C_0| [\nabla f]\le 1\}}\int \mu_N f
\]
On the other hand, for any function $f$ with compact support we have as a well known consequence of Fubini's Theorem (see for instance \cite{lieb}, where this is named the bath-tub principle)
\[
\int \mu_N f_+=\int_0^{+\infty}\(\int_{\{f>t\}}\mu_N \) dt
\]
and
\[
\int \mu_N f_-=\int_0^{+\infty}\Big(\int_{\{f<-t\}}\mu_N \Big) dt. 
\]
Together with \eqref{maxmingrad}, this proves the result.\cqfd

\begin{lem}
\label{auxilary1}
Assuming Property~I of Proposition~\ref{procritical}, there exists $C>0$ such that for any  integer $N$ and any  finite  $B\subset \Z^2$, we have
\[
\mu_N(B)\le 4CN\, \per(B)
\]
\end{lem}
\pr
Let $B_1,\cdots, B_k$ be the connected components of $B$. Then  we have  disjoint unions
$B=\bigcup_{i=1}^k B_i$ and  $\partial B=\bigcup_{i=1}^k \partial B_i$. Set $\tilde B_i:=\bigcup_{p\in B_i} K_p^N$.  We have $\mu_N(B_i)=\sharp\(\tilde B_i\cap A\),$ hence
\[
\mu_N(B_i)\le C\diam(\tilde B_i)
\]
Now assume $\tilde p=(\tilde p_1,\tilde p_2)$ and $\tilde q=(\tilde q_1,\tilde q_2)$ are in $\tilde B_i$ and such that $ \diam(\tilde B_i)=\|\tilde p-\tilde q\|$. Without loss generality, we may assume that $\|\tilde p-\tilde q\|\le 2(\tilde p_1-\tilde q_1)$.  There exists $p=(p_1,p_2)$ and $q=(q_1,q_2)$ in  $B_i$ such that  $\tilde p\in K_p^N$ and $\tilde q\in K_q^N$. Moreover, 
$$
\tilde p_1-\tilde q_1 =  N(p_1-q_1) + (N-1)\le N(p_1 - q_1 +1).
$$
On the other hand, from the connectedness of $B_i$, for any integer $x\in[p_1,q_1]$ we have $B_i\cap\{r\}\times\Z\neq\varnothing$ hence writing $m_x = \min\{y\mid(x,y)\in B_i\}$ and $M_x = \max\{y\mid(x,y)\in B_i\}$, the two edges  $\((x,m_x),(x,m_x-1)\)$ and $\((x,M_x),(x,M_x+1)\)$ belong to $\partial B_i$. It follows that 
$$
 \per(B_i) = \sharp\partial B_i \ge 2 (p_1-q_1+1),
$$
and then 
$$
\mu_N(B_i)\le C N \per(B_i), \quad 
\mu_N(B)=\sum_i \mu_N(B_i)\le CN \sum_i \per(B_i)= C N \per(B).
$$
\cqfd

As a consequence, we obtain

\begin{cor}
Assuming Property~I of Proposition~\ref{procritical}, there exists $C>0$ and for any integer $N>1$  there exists a $1$-form $\varphi$ such that 
\beq\label{div}
-\div(\varphi)=\mu_N
\eeq
and for every edge $a\in{\cal A}$,
\beq\label{cn}
|\varphi(a)|\le C N.
\eeq
\end{cor}
\pr It follows from Lemmas \ref{Coarea} and \ref{auxilary1} that
\[
\int_0^{+\infty}\mu_N(\{f>t\})\le C N \int_0^{+\infty}\per(\{f>t\})=CN[\nabla f_+]
\]
and 
\[
\int_0^{+\infty}\mu_N(\{f<-t\})\le CN \int_0^{+\infty}\per(\{f>t\})=CN[\nabla f_-].
\]
This implies using Lemma~\ref{express} that 
\[
\alpha\le CN \max_{\nabla f\in C_0, [\nabla f]\le 1}[\nabla f] = CN. 
\]
Using the definition of $\alpha$, there exists a $1$-form $\varphi$ with the desired properties (changing the constant to $2C$ for instance).\cqfd

\pr[ Proof of Proposition \ref{procritical}.] We construct the bijective map $\Phi:\Lambda\to \Z^2$. This is done by specifying the for every $p,q\in\Z^2$ the number of points in $\Lambda\cap K_p^N$ whose images by $\Phi$ belong to  $\Z^2\cap K_q^N$, as follows:
 \[
n_{p\to q}:=\left\{
\begin{array}{lll}
\max (\varphi(p,q),0)&\mbox{ if }(p,q)\in{\cal A}\\
\ds \sharp(\Lambda\cap K_p^N)-\sum_{(p,q)\in {\cal A}} n_{p\to q}&\mbox{ if }p=q\\
0&\mbox{ otherwise,}
\end{array}
\right.
\]
where $\varphi$ is a flow satisfying \eqref{div}, \eqref{cn}.

Now, for the numbers $n_{p\to q}$ to indeed correspond to a bijective map $\Phi$ we need to check some of their properties.
\medskip

\noindent{\em Property  1.} If $N$ is chosen large enough, then for any $p,q\in\Z^2$, we have $n_{p\to q}\ge 0$. This is clear when $p\neq q$.  In the case $p=q$, we note that there are exactly  4 edges coming out of  $p$. Thus, from \eqref{cn}
and the fact that $\sharp(\Lambda\cap K_p^N)\ge N^2 - CN$ we find (with another constant $C$ still independent of $N$).
$$
n_{p\to p}\ge N^2- CN.
$$
Thus we may indeed choose  $N$ large enough so that indeed $n_{p\to q}\ge 0$ for any $p,q\in\Z^2$.\medskip

\noindent{\em Property  2.} This one is clear from the definition of $n_{p\to q}$: For any $p\in\Z^2$ we have 
$$
\sum_q n_{p\to q} = \sharp(\Lambda\cap K_p^N).
$$\medskip

\noindent{\em Property  3.} For any $q\in\Z^2$ we have
$$
\sum_p n_{p\to q} = N^2.
$$
Indeed, fixing $q\in\Z^2$ and all the sums below being with respect to $p$,
\begin{align*}\sum_p n_{p\to q} &=  n_{q\to q}+ \sum_{(p,q)\in{\cal A}} n_{p\to q} \\ &= \sharp(\Lambda\cap K_q^N)-\sum_{(q,p)\in {\cal A}} n_{q\to p}+ \sum_{(p,q)\in{\cal A}} n_{p\to q}\\
&=  \sharp(\Lambda\cap K_q^N)+\sum_{(p,q)\in {\cal A}, \varphi(p,q)\ge 0} \varphi(p,q) - \sum_{(q,p)\in{\cal A},\varphi(q,p)\ge 0} \varphi(q,p).
\end{align*}
Now since $\varphi(p,q) = -\varphi(q,p)$ we have 
$$ \sum_{(p,q)\in {\cal A}, \varphi(p,q)\ge 0} \varphi(p,q) - \sum_{(q,p)\in{\cal A},\varphi(q,p)\ge 0}\varphi(q,p) = \sum_{(p,q)\in {\cal A}} \varphi(p,q) =- \div\varphi(q).$$
Using \eqref{div} this sum is equal to $\mu_N(q) = N^2 - \sharp(\Lambda\cap K_q^N)$, hence $\sum_p n_{p\to q} = N^2$.

The three properties insure that there exists a bijection $\Phi:\Lambda\to\Z^2$ such that for any $p,q\in\Z^2$ we have 
$$ n_{p\to q} = \sharp\{x\in\Lambda\cap K^N_p\mid \Phi(x)\in \Z^2\cap K^N_q\}.$$
Since $n_{p\to q}\neq 0$ implies $\|p-q\|\le 1$, we find that $\|\Phi(x) - x\|\le 2\diam(K^N)$, for any $x\in\Lambda$.
\cqfd

\begin{rem} The conclusion of Theorem~\ref{critical} holds under the following, less restrictive assumption on $\Lambda$, which is assumed to be uniform, but not necessarily a subset of $\Z^2$:
\begin{itemize}
\item[i)] There exists some positive constant $C>0 $ such that for any $x\in\R^2$ and any $R>1$, one has
$  |\sharp(\Lambda\cap B(x,R))-\pi R^2|\le C R. $
 \item[ii)] There exists some positive integer $N_0\in \N$  such that for any $p\in\Z^2$, one has 
$ \sharp\(K^{N_0}_p\cap \Lambda\)\le N_0^2.$
\end{itemize}
Indeed, the second assumption, implies that there exists an  injective map 
$$
\Phi_p: K^{N_0}_p\cap \Lambda\to K^{N_0}_p\cap \Z^2.
$$
We define $\Phi:\Lambda\to\Z^2$ to be the injective map whose restriction to $K^{N_0}_p$ is $\Phi_p$ for any $p\in\Z^2$ and let $\Lambda_1=\Phi(\Lambda)$. Then $\Lambda_1$ is of the form $\Z^2\setminus A$, with $A$ satisfying \eqref{disc}.   Theorem~\ref{critical} implies that $\tilde W(\Lambda_1) < +\infty$ and then from \eqref{pro-move} we deduce that $\tilde W(\Lambda)<+\infty$. 
\end{rem}

We conclude this section with 

\begin{thm3bis}
\label{criticalbis}
Let $\Lambda\subset \R^2$ be  discrete and uniform, and of the form $\Lambda = \Lambda_1\times\Z$, where $\Lambda_1\subset\R$. 

If there exists $C>0 $ such that for any $x\in\R^2$ and  $R>1$ we have $|\sharp (\Lambda\cap K(x,R))- R^2|\le C R$ --- where $K(x,R)$ is the square with sidelength $R$ and center $x$ ---  then $\tilde W(\Lambda)<+\infty$.
\end{thm3bis}

\pr The proof of the theorem will follow the same strategy as for Theorem~\ref{critical}, except that we work now in one dimension. For any integer $N>0$ and $p\in\Z$ we let $I^N_p = [pN,(p+1)N)$ and  $\mu^N(p) = N- \sharp(\Lambda_1\cap I^N_p)$. We consider the graph with $\Z$ as the set of vertices and the set of edges
$$\cal A = \{(p,q)\mid p,q\in\Z, |p-q| = 1\}.$$

We claim that there exists $C>0$, and for any integer $N>0$  a $1$-form $\varphi:\cal A\to\R$ such that 
\beq\label{flot1d} -\div(\varphi)=\mu_N,\quad \|\varphi\|_\infty\le C.\eeq
Indeed we define $\varphi$ as follows:
 \[
\varphi((k,k+1))= \left\{
\begin{array}{ll}
\ds 0 &\mbox{ if }k = 0 \\
\ds - \sum_{i=1}^k \mu_N(i)&\mbox{ if }k\ge 1\\
\ds   \sum_{i=k+1}^0 \mu_N(i)&\mbox{ if }k< 0.
\end{array}
\right.
\]
It is clear that $-\div(\varphi)= \mu_N$. Moreover, for instance if $k\ge 1$, then 
$$\varphi((k,k+1)) = -\sum_{i=1}^k \(N- \sharp(\Lambda_1\cap I^N_p)\) = \sharp(\Lambda_1\cap [N, (k+1)N))- kN.$$
But considering the square $K = [N, (k+1)N)\times[N, (k+1)N)$, we have 
$$k N - \sharp \(\Lambda_1\cap [N, (k+1)N)\) = \frac1{kN}\( (kN)^2 -  \sharp (\Lambda\cap K)\),$$ 
and thus using the hypothesis of the theorem  we deduce that  $|\varphi((k,k+1))|\le C$ as claimed. 

Now we choose  $N\ge 2C+1$ and following the proof of Proposition~\ref{procritical} we can construct  a bijective map $\Phi: \Lambda_1\to \Z^2$ such that $|\Phi(p) - p|$ is bounded independently of $p$. This induces a bijection with the same property from $\Lambda$ to $\Z^2$, which proves Theorem~\ref{criticalbis}, using Proposition~\ref{pro-move}. 
\cqfd

\section{A Penrose lattice}\label{penrose}
We now describe the construction of a Penrose-type lattice $\Lambda$ such that $\tilde W(\Lambda)<+\infty$. Of course it would be better to show that $\Lambda$ satisfies the hypothesis of Theorem~\ref{thm2}, but this to our knowledge an open problem. 

For the simplicity, we consider the Robinson triangle decompositions in the Penrose's second tilling (P2)--kite and dart tiling, or in the Penrose's third tilling (P3)--rhombus tiling, (for the reference see \cite{Se}). The construction is as follows: $\Omega_1$ and $\Omega_2$ are  two Robinson triangles, namely,  $ \Omega_1$ is an acute Robinson triangle having  side lengths $(1,1, \varphi)$, while $ \Omega_2$ is obtuse one with sidelengths $(\varphi, \varphi, 1)$,
where $\varphi=(1+\sqrt{5})/2$;  the scaled-up domain $\varphi\Omega_1$ decomposes as the union of a copy of  $\Omega_1$ and a copy of $\Omega_2$, where the interiors are disjoint --- and such that $\varphi\Omega_2$ decomposes as the union of
one copy of $\Omega_1$ and two copies of $\Omega_2$ with disjoint interiors (see figure).

\begin{figure}
\begin{center}
\resizebox{0.8 \linewidth}{!}{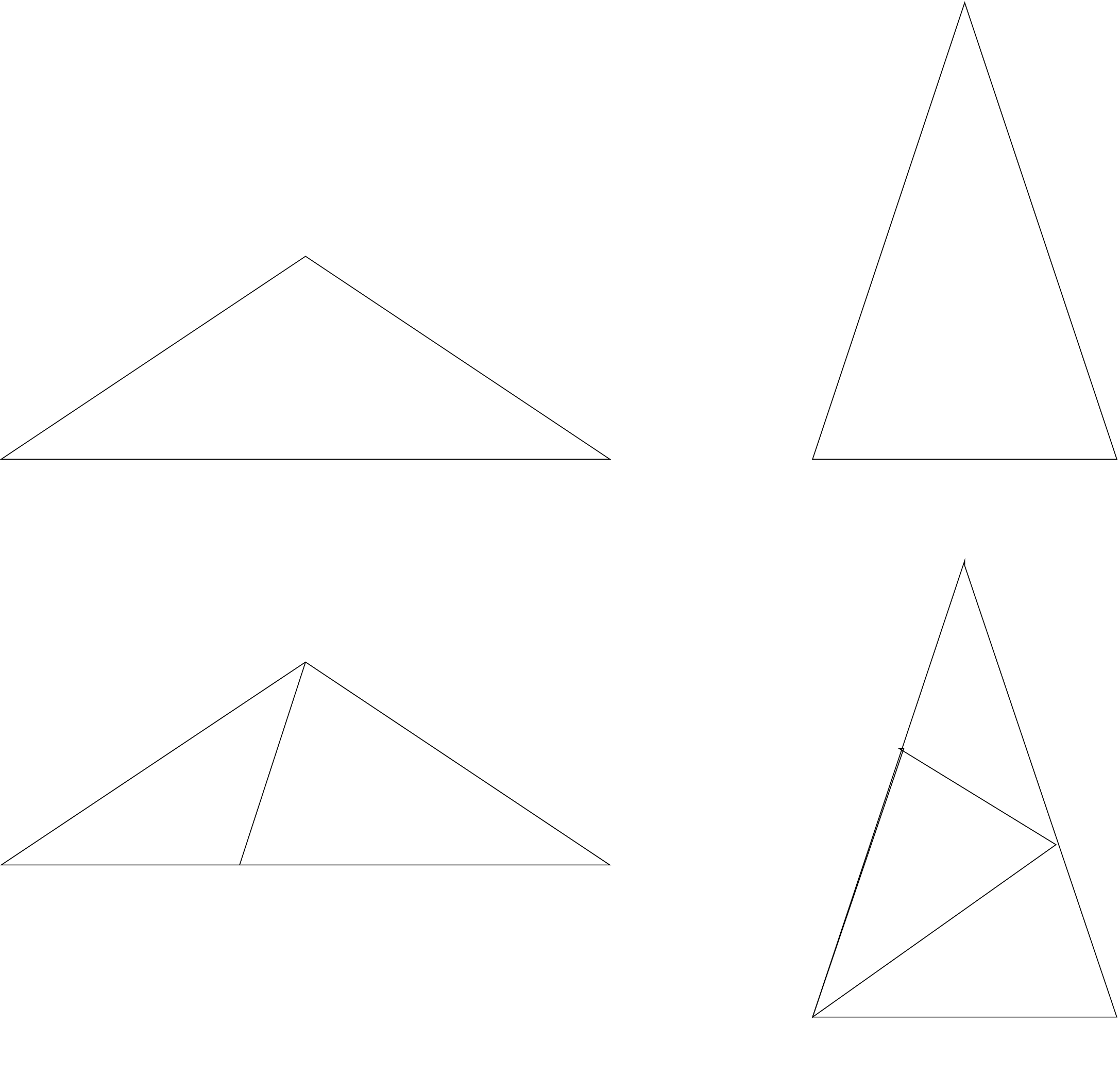}
\caption{}
\end{center}
\end{figure}

For $i=1,2$ we  choose a point $p_i$ in the interior of $\Omega_i$. 

Then we proceed by induction, starting with $\Omega_1$ choosing $p_1$ as the origin, then scaling up  by $\varphi$, then decomposing, then scaling up again, then decomposing each piece, etc... After $n$ steps we have a (large domain) $\varphi^n \Omega_1$ decomposed a number of copies of either $\Omega_1$ or $\Omega_2$. In each copy we have a distinguished point, the union of which is denoted $\Lambda_n$. As $n\to+\infty$ and modulo a subsequence, $\Lambda_n$ converges to a discrete set $\Lambda$, which is uniform since the distance between two point is no less than $\min\(d(p_1,\partial\O_1), d(p_2,\partial\O_2)\). $

\begin{thm}
We have $\tilde W(\Lambda)<+\infty$.
\end{thm}

\pr For each $n$ we define a current $j_n$ as follows. On each copy of $\O_i$ we let $j_n$ be equal to (a copy of) $\nabla U_i$, where 
\[
\left\{
\begin{array}{lllll}
-\triangle U_i &=&\delta_{p_i}-\frac{1}{|\O_i|} &\mbox{ in }\Omega_i\\
\frac{\partial U_i }{\partial\nu}&=&0& \mbox{ on }\partial\Omega_i.
\end{array}
\right.
\]
Then $j_n$ converges as $n\to +\infty$ to a current $j$ such that the following holds in $\R^2$
$$-\div( j )=\sum_{p\in\Lambda} \delta_{p}- \alpha,$$
where $\alpha = 1/|\O_i|$ on each copy of $\O_i$. It is not difficult to check that $W(j)<+\infty$, but the background density $\alpha$ is not constant. We need to add a correction to $j$, which is the object of the following
\begin{lem}
\label{lem4.6}
There exist $m\in\R$ and a solution of the following equation in $\R^2$ 
\beq\label{jprime}-\div( j') =\alpha - m\eeq
such that $\|j'\|_{\infty}<+\infty,$
\end{lem}

Assuming the lemma is true we let $\tilde j = j + j'$. Then $-\div (\tilde j) = \sum_{p\in\Lambda} \delta_{p}- m$ thus $\tilde j\in\Fl$ for the background $m$, and the fact that $W(j)<+\infty$ and $j'\in L^\infty$ implies that $W(\tilde j)<+\infty$ and the Theorem. 
\cqfd

\pr[Proof of Lemma~\ref{lem4.6}] The current $j'$ is obtained as the limit of $j_n$, where $j_n$ solves 
\beq\label{jn}
\left\{
\begin{array}{lllll}
-\div (j_n) &=&\alpha_n - m_n&\mbox{ in
}\varphi^n \Omega_1\\
j_n\cdot\nu &=&0& \mbox{ on
}\partial(\varphi^n\Omega_1),
\end{array}
\right.
\eeq
where $\alpha_n:\varphi^n\O_1\to \R$ is the function equal to $1/|\O_i|$ on each of the copies of $\O_i$, $i=1,2$ which tile $\varphi^n\Omega_1$, and where $m_n$ is equal to the average of $\alpha_n$ on $\varphi^n \O_1$. 

The current $j_n$ is defined recursively. First we define the equivalent of $\alpha_n$ for $\O_2$-type domains: For any integer $n$ we tile $\varphi^n\O_2$ by one copy of $\varphi^{n-1}\O_1$ and two copies of $\varphi^{n-1}\O_2$, then we tile each of the three pieces, etc... until we have tiled $\varphi^n\O_2$ by copies of either $\O_1$ or $\O_2$. then we let $\beta_n:\varphi^n\O_2\to \R$ be the function equal to $1/|\O_i|$ on each of the copies of $\O_i$, $i=1,2$. We also define $q_n$ to be the equivalent of $m_n$, i.e. the average of $\beta_n$ on $\varphi^n\O_2$. Finally we define $\bar \jmath_n$ to be the equivalent of $j_n$ for type $2$ domains, i.e. the solution of \eqref{jn} with $\alpha_n$ replaced by $\beta_n$, $m_n$ replaced by $q_n$ and $\O_1$ replaced by $\O_2$. 

Below it will be convenient to  abuse notation by writing $\varphi^n\O_i$ for a copy of $\varphi^n\O_i$. Then we have $\varphi^n\Omega_1=\varphi^{n-1}\Omega_1\cup\varphi^{n-1}\Omega_2.$ We let 
\beq \label{recu} j_n = j_{n-1} \indic_{\varphi^{n-1}\Omega_1}+ \bar\jmath_{n-1} \indic_{\varphi^{n-1}\Omega_2}+\nabla U_n\indic_{\varphi^n\Omega_1},\eeq 
where
\beq\label{un}
\left\{
\begin{array}{lllll}
-\triangle U_n &=&  (m_n - m_{n-1})\indic_{\varphi^{n-1}\Omega_1}+(m_n - q_{n-1})\indic_{\varphi^{n-1}\Omega_2}&\mbox{ in
}\varphi^n \O_1\\
\frac{\partial U_n }{\partial\nu}&=&0& \mbox{ on
}\partial(\varphi^n \O_1).
\end{array}
\right.
\eeq
It is straightforward to check that $j_n$ satisfies \eqref{jn} assuming $j_{n-1}$ and $\bar\jmath_{n-1}$ do. 

The relation \eqref{recu} is the recursion relation which repeated $n$ times allows to write $j_n$ as equal to a sum of on the one hand error terms $\nabla U_k$ (or their type $2$ equivalent that we denote $V_k$), for $k$ between $1$ and $n$, and on the other hand of a vector field which on each elementary tile of type $\O_1$ of $\varphi^n\O_1$ is equal to $j_0$ and on a tile of type $\O_2$ is equal to $\bar\jmath_0$. However from \eqref{jn} we may take $j_0 = 0$ and $\bar\jmath_0 = 0$, thus we are left with evaluating the error terms.

{\bf Claim:} There exists $C>0$ such that for any integer $k>0$ we have 
$$ \|\nabla U_k\|_\infty, \|\nabla V_k\|_\infty \le C \varphi^{-3k}.$$
This clearly proves that the sum of errors for $k=1\dots n$ is bounded in $L^\infty$ independently of $n$ and therefore that  $\{j_n\}$ is bounded in $L^\infty$. Then the limit $j'$ is in $L^\infty$.

To prove the lemma, it remains to prove the claim, and to show that $j'$ satisfies \eqref{jprime} for some $m\in\R$, which in view of \eqref{jn} amounts to showing that $\{m_n\}_n$ converges. For this we define $u_{2n}$ (resp. $u_{2n+1}$) be the number of elementary tiles of type $\Omega_1$ (resp. $\Omega_2$) in $\varphi^n\O_1$. We define similarly $v_{2n}$ and $v_{2n+1}$ by replacing $\O_1$ by $\O_2$. Therefore  $u_0=1$, $u_1=0$, $v_0 = 0$, $v_1 = 1$. We have the following recurrence relations
$$ u_{2n+2} = u_{2n} + u_{2n+1},\quad u_{2n+3} = u_{2n} + 2 u_{2n+1},$$
which we can summarize as the single relation $u_{n+2}=u_{n+1}+u_n$. Similarly $v_{n+2}=v_{n+1}+v_n$. It follows that
$$u_n =\varphi^n \frac{1}{\varphi+2} + (-\varphi)^{-n}\frac{\varphi +1}{\varphi+2} ,\quad v_n ={\varphi^n} \frac\varphi{\varphi+2} + (-\varphi)^{-n}\frac{-\varphi}{\varphi+2}. $$
We have $u_n =  a \varphi^n + O(\varphi^{-n})$ and $v_n =  b \varphi^n + O(\varphi^{-n})$ with $a= \frac{1}{\varphi+2}$ and $b= \frac\varphi{\varphi+2}$ strictly positive. Then we easily deduce that 
$$m_n = \frac{u_{2n} + u_{2n+1}}{u_{2n}|\O_1| + u_{2n+1}|\O_2|} = m + O(\varphi^{-4n}),$$
where 
$$ m = \frac{1 + \varphi}{|\O_1| + \varphi|\O_2|},$$
and similarly that $q_n = m + O(\varphi^{-4n}).$ This proves in particular the convergence of $\{m_n\}_n$. Moreover it shows that the right-hand side of \eqref{un} is bounded by $C\varphi^{-4n}$. By elliptic regularity (lemma \ref{lem3.1} and lemma \ref{lem2}) we deduce that 
$$\|\nabla U_n\|_\infty\le C |\varphi^n\O_1|^{\frac12} \varphi^{-4n} = C |\O_1|^{\frac12} \varphi^{-3n},$$
and a similar bound for $V_n$. This proves the claim, and the lemma
\cqfd

\begin{rem} The above construction could easily be generalized to similar recursive constructions.
\end{rem}

\noindent{\bf Acknowledgments.} {The authors wish to thank Y.Meyer for helpful discussions.}

\end{document}